\documentclass{aastex631}

\usepackage{comment,amsmath}

\begin{document}

\title{Evaluating the $\Sigma$-effect Model of the Solar Hemispherical Helicity Bias via Direct Numerical Simulations}

\correspondingauthor{Nicholas Brummell}
\email{brummell@ucsc.edu}

\author{Jacob B. Noone Wade}
\affiliation{Department of Applied Mathematics, Baskin School of Engineering, University of Califonia, Santa Cruz \\
1156 High Street \\
Santa Cruz, CA 95064}

\author{Nicholas H. Brummell}
\affiliation{Department of Applied Mathematics, Baskin School of Engineering, University of Califonia, Santa Cruz \\
1156 High Street \\
Santa Cruz, CA 95064}



\let\manuscript=y

\begin{abstract}
The Solar Hemispherical Helicity Rule(s) (SHHR) is a term used to represent a bias observed in proxies for the magnetic helicity in active regions at the solar surface.  The SHHR states that  predominantly negative magnetic helicity is observed in active regions in the northern hemisphere, whereas predominantly positive is found in the southern.
The $\Sigma$-effect model of \cite{longcope1998flux} is one of the most cited models for the explanation of the SHHR.  In this model, the magnetic structures derive the bias in their magnetic helicity from  the kinetic helicity of the turbulent convection through which they travel, where the latter is handed owing to the rotational influence of the star.  The original paper built an elegant mathematical model for the dynamics of thin flux tubes influenced by parameterized helical turbulence.  Here, we attempt to explore the conceptual ideas of this original simplified model using fully-nonlinear, three-dimensional, Cartesian-domain  simulations of isolated, finite cross-sectional, twisted magnetic flux structures rising though rotating, overshooting, turbulent compressible convection.  We look for evidence of a correlation between the kinetic helicity content of the turbulence and the evolving magnetic helicity of the structures.  We find little evidence of such a relationship, and do not even find any clear hemispheric dependence.  Although these simulations are far from a perfect representation of the ideas, this work raises many questions about the potential efficacy of the $\Sigma$-effect in reality.

\end{abstract}



\section{Introduction} \label{sec:intro}

Active regions (ARs) appearing at the solar photosphere are thought to be the manifestation of emerging magnetic flux that buoyantly rises from the deep solar interior. The observations of the nature of ARs and their sunspots leads to the common notion that the emerging magnetic flux is organized into coherent cylindrical concentrations of predominantly toroidal but twisted magnetic field, often referred to colloquially as magnetic ``flux tubes". Modelling has shown \citep[see e.g.][]{schussler1979magnetic, emonet1998physics} that for a flux tube to survive transit to the surface from a presumed point of origin deep in or below the solar convection zone, even in the absence of convection, then the magnetic field lines of the flux tube must be sufficiently twisted to provide a cohesive tension force to prevent the vortex wake of the structure from ripping itself apart. The field line topology of the flux tube is therefore already clearly an essential component of its dynamics.

Magnetic helicity is the quantity that measures field line topology and complexity, encapsulating and binding together the concepts of twist, writhe and linkage of magnetic fields \citep[see e.g.][]{berger1984topological}. The twist is the winding of the field lines around an axis, writhe is the winding of the axis itself, and linkage quantifies the knottedness of fieldlines.  Magnetic helicity is defined as $H_m = \int_V \mathbf{A} \cdot \mathbf{B} ~dV$, where $\mathbf{A}$ is the vector potential corresponding to the magnetic field, $\mathbf{B}=\nabla \times \mathbf{A}$ defined over a fluid volume $V$. Magnetic helicity holds great significance for various reasons: (a) it is a conserved quantity in ideal MHD and thus is a strong constraint in solar dynamo theory \citep[see e.g.][]{Field:1986}; (b) its accumulation in ARs allows them to store a significant amount of magnetic energy, which, when released via non-ideal processes like magnetic reconnection (topological rearrangement of field lines) is thought to be related to eruptive events on the solar atmosphere above \citep[see e.g.][]{Rust:Kumar:1996, Low:1996}. While the amount of magnetic helicity in ARs is related to their magnetic energy content, its sign is an indicator of the handedness (chirality) of the winding of field lines. Interestingly, in solar observations, the sign of various proxies for elements of the magnetic helicity show a consistent and significant temporal coherency, forming the basis of the so-called ``solar hemispherical helicity rule(s)'' (SHHR) discussed in more detail below. Magnetic helicity, however, is unfortunately not directly observable because the vector potential, $\mathbf{A}$, is not an measurable quantity, and it's gauge invariance make its derivation from observed quantities non-unique. This limitation is typically overcome by measuring $H_m$ with respect to a potential magnetic field, creating a quantity known as known as relative magnetic helicity \citep{berger1984topological}, or by using alternative measures of magnetic field that can be directly observed as proxies, as described next. 

Current helicity is a related quantity that is more easily observable. It is defined as $H_c = \int_V \mathbf{B} \cdot \mathbf{J}~dV$, where $\mathbf{J}=\nabla \times \mathbf{B}$ is the electric current density. Qualitatively, this measure also represents the complex topological features of magnetic field lines in the ARs.  
Current helicity is not strictly conserved in ideal MHD though. However, it has been shown through observations and theory that the sign of current helicity is commensurate with the sign of magnetic helicity in ARs \citep{seehafer1990,Gosain_Brandenburg_2019}. Elements of current helicity have therefore been used extensively as proxies for the actual magnetic helicity in studying the helical content of solar surface features. In general, only the surface-normal component of the total current helicity, relating $J_z=(\nabla \times \mathbf{B})_z$ and the vertical component of $\mathbf{B}$, $B_z$, is computed from observations of the surface field \citep[see][]{pevtsov2003}. When visualized in the context of an idealized cylindrical ``flux tube" emerging vertically through the surface (that is, imagining the ``legs'' of an $\Omega$-shaped magnetic flux structure as they pass through the solar surface), the component observed corresponds to the magnetic twist of the flux tube, although it is often referred to generically as helicity in this context. 

Many observations of vector magnetic fields within different solar surface magnetic features (ARs, sunspots, penumbrae, filaments, coronal sigmoids, etc) have been analyzed for such helicity proxies \citep[for an overview, see][]{pevtsov2003}. They all reveal a hemispheric helicity rule, where there is a tendency for the photospheric magnetic fields to have negative helicity in the northern hemisphere and positive helicity in the southern hemisphere. 
 An interesting feature of these observations is that these rules persist from one half of the full 22 year solar cycle to the next,
in contrast with other well-established rules, such as Hale's polarity law \citep{hale1919magnetic}, where the polarities of the leading and trailing sunspots flip between halves of the full cycle. However, the agreement with the sign rule is fairly weak, meaning that only about $60-80\%$ of ARs follow the  rule. Observations also show varying levels of agreement in time \citep{Hagino:Sakurai:2005}, with some observations indicating stronger disagreement with the rule towards the end of a solar cycle \citep{Hao:Zhang:2011}. All of these helicity observations combined fall collectively under the umbrella terminology of ``the SHHR''.

Many models seek to explain the existence of the SHHR, but possibly the most cited theory currently is that of \cite{longcope1998flux}, known colloquially as ``the $\Sigma$-effect". This paper describes an elegant mathematical model based on the evolution of a thin flux tube model. Thin flux tube models \citep[see e.g.][]{Defouw:1976, Spruit:1981} simply replace real finite cross-sectional flux tubes by a line along which various forces can act and qualities can be described; in particular, magnetic buoyancy and drag forces act and a twist can be ascribed. \cite{longcope1998flux} extends earlier thin flux tube models to include an explicit equation for the evolution of the twist of the tube, derived in \cite{Longcope:Klapper:1997}.  Twist can then change along the thin flux tube axis depending on the stretching and rotation of the axis line, and also via a term, denoted by a $\Sigma$ in the equation in the paper (hence the name of the effect), that describes the production of twist from the buffeting of the thin flux tube axis by the surrounding convective turbulence.  If the turbulence were isotropic and homogeneous, this buffeting might be expected to be random, and therefore, likewise the writhe and twist of the thin flux tube would be random.  But if the convective turbulence were to be biased, as is expected if it is significantly influenced by the rotation of the star, then \cite{longcope1998flux} argued that the biased motions would induce a specific net chirality in the writhe of the thin flux tube axis, which would then in turn, assuming magnetic helicity is conserved, induce a twist of the opposite handedness to compensate.  There is therefore a relationship between the kinetic helicity of the turbulence and the measured magnetic helicity (really the twist, since the writhe component is not measured) of the emerging thin flux tubes in the model.  Assuming that rising fluid elements generally expand in the compressible convective turbulence of the solar convection zone, then they should spin down under the action of any significant Coriolis force, leading to an expectation of generally negative kinetic helicity (left-handed; assuming radially outward and prograde motions are positively valued) in the northern hemisphere, for example.  These motions would induce a positive (right-handed) writhe in the magnetic axis of the thin flux tube (lifting a loop out of the originally-horizontal axis and turning it counter-rotation), and thereby a negative (left-handed) twist for the northern hemisphere.  All chiralities are reversed in the southern hemisphere, of course, leading to positive twist in an emerging tube there.  These handedness rules are indeed commensurate with what is observed in the SHHR.  Coupled with a chosen model for helical turbulence, \cite{longcope1998flux}'s inciteful model demonstrated that sufficient twist of the correct sign could be generated to explain the SHHR observations.

The $\Sigma$-effect is one plausible explanation of how an {\sl isolated} flux tube {\sl acquires} the appropriate twist for the SHHR on its rise towards emergence.  Other models have adopted similar, acquisition-based ideas.  For example, the acquisition of twist may be attributed to helicity conservation after the direct writhing of a rising flux tube (due to the action of the Coriolis force on the flows in a rising $\Omega$ loop), rather than indirect writhing via the rotationally-influenced convective turbulence \citep{Wang:Sheeley:1991,Dsilva:Choudhuri:1993, Fan:Fisher:Mcclymont:1994,Wang:2013}.  Since the large-scale writhe is perhaps measurable via Joy's Law, the related twist content can be surmised.  However, this process is not capable of providing sufficient twist to match the observations \citep{Longcope:Linton:Pevtsov:Klapper:1999,Fan:Gong:2000}.  Furthermore, the expected latitudinal dependence from this idea has not been observed \citep{Pevtsov:Canfield:1999,Holder:Canfield:Mcmullen:Nandy:Howard:Pevtsov:2004}.
\cite{Choudhuri:2003} suggested a different acquisition-based model where an untwisted tube rises into a near-surface large-scale background poloidal field arising as a result of the breakup of tilted active regions.   \citet{Chatterjee:Choudhuri:Petrovay:2006} and \citet{Hotta:Yokoyama:2012}, using models and simulations respectively,   subsequently studied the accretion of twist onto an initially untwisted flux tube that is rising into specific formulations of background field. However, most of these acquisition-based models either struggle to achieve sufficient twist or adhere to the SHHR strictly, thereby being in discord with other elements of the observations.  Only the $\Sigma$-effect seems to provide reasonable results in this ``acquisition'' class of models.

Another more recent model, examined in a series of papers \citep{Manek:Brummell:Lee:2018, Manek:Brummell:2021, Manek:Pontin:Brummell:2022, ManekBrummell3D}, does perhaps show some promise.  These numerical models follow finite cross-section flux concentrations evolving within a larger-scale horizontal background field (both with and without convective turbulence).  This model works very differently from the acquisition-based models.  The new dynamics involve a filter mechanism that selects magnetic structures with the ``correct'' characteristics (i.e. commensurate with the SHHR) to emerge preferentially over magnetic structures that do not match the SHHR.  This model has some distinctly positive aspects (such as the natural production of a weak adherence to the rule, temporal variation over the cycle, independence of latitude etc) and could, in principle, operate in tandem with any of the other effects.  However, our purpose with this paper is not to examine this newer mechanism further, but instead to test out the efficacy of the $\Sigma$-effect in a more realistic setting than the elegant but simple model of \cite{longcope1998flux}.

The series of studies cited above are direct numerical simulations (in up to three dimensions) of finite cross-sectional flux tubes and follow the full dynamical evolution of the field and its topology, unlike the model of \cite{longcope1998flux} that more simply solves the heuristic thin flux tube equations.  
To our knowledge, the $\Sigma$-effect has, to date, never been tested by similar more realistic simulations, and the pursuit of this is our intention here. We therefore perform 3D MHD simulations of the evolution of initially-horizontal, twisted, finite cross-section flux tubes in the presence of fully resolved rotating compressible convection. Here, unlike the model of \cite{longcope1998flux}, we must assume some initial twist in the magnetic structure to allow a coherent rise, otherwise breakup of the magnetic structure ensues even in the absence of convection \citep[see e.g.][]{emonet1998physics}.  We examine how the helical properties of the flux tubes evolve during transit from an underlying radiative zone through an overlying convection zone, for varying levels of rotational influence on the convective turbulence.  If the ideas of the $\Sigma$-effect are robust, then there should be some correlation between the rotational influence on the convection and the deviation of the writhe and the twist from their initial levels during the flux tube's evolution.  Since we model fully-resolved 3D flux tubes, we have the ability to follow the actual field lines of the flux tube and thereby to calculate helical measures directly as a function of time without any assumptions.

This paper is arranged as follows.  In Section \ref{sec:model}, we describe our 3D numerical setup in detail.  In Section 
\ref{sec:prelims}, we describe our choices of parameter values, the measures we will use, and the evolution of the model when there is no convection present.  In Section \ref{sec:results}, we describe the findings of running this model in the presence of convection for various degrees of rotational influence and various levels of turbulence.  In Section \ref{sec:disc} we discuss these findings and their implications for theories of the SHHR.

\section{Model Formulation}\label{sec:model}

\newcommand{\sig}{{Pr}}
\newcommand{\kap}{K}
\newcommand{\U}{\mathbf{u}}
\newcommand{\B}{\mathbf{B}}

\subsection{Two-layer overshooting convection model}
\label{sec:twolayer}

Our model examines the evolution of magnetic fields in a domain of turbulent
overshooting compressible convection.
The Cartesian three-dimensional, two-layer overshooting convection numerical model used in this paper is essentially the same as that used in previous papers, such as 
\cite{brummell2002penetration} and
\cite{Tobias:etal:2001}, but with different magnetic field initial conditions.
We utilise a rectilinear domain 
containing a fully compressible but
ideal gas confined between two horizontal, impenetrable, stress-free
boundaries.  The Cartesian box spans
    ${0 \le \tilde x \le x_{\rm m}d}$ 
and ${0 \le \tilde y \le y_{\rm m}d}$ in the horizontal
and ${0 \le \tilde z \le z_{\rm m}d}$ in the vertical, 
with the $\tilde z$-axis pointing downwards, and where $d$ is the depth of the upper convecting portion of the domain.
The upper surface is held at a fixed
temperature $T_0$ whereas a constant vertical temperature gradient $\Delta$ is
maintained at the lower boundary.  The fields are assumed to be periodic
in the two horizontal directions.  The specific heats $c_p$ and $c_v$,
shear or dynamic viscosity $\mu$ (related to the kinematic viscosity $\nu=\mu/\rho$,
where $\rho$ is the density) and the gravitational
acceleration $g$ are assumed constant.  

In a single layer of depth $d$ ($z_m=1$) and with constant thermal
conductivity $K$, the
temperature $T_p$, density $\rho_p$ and pressure $p_p$ can exist in 
hydrostatic balance in a polytropic state
\begin{mathletters}
\begin{subequations}   
\label{eqn_polytrope}
\begin{align}
T_p/T_0
&=
(1 + \theta \tilde z/d),
\\
\rho_p/\rho_0
&=
(1 + \theta \tilde z/d)^m,
\\
p_p/p_0
&=
(1 + \theta \tilde z/d)^{m+1},
\end{align}
\end{subequations}
\end{mathletters}
\noindent where $\rho_0$ is the density at the upper boundary, $p_0 = (c_p -
c_v)T_0 \rho_0$, $m = -1 +g/\Delta (c_p - c_v)$ is the polytropic index
and $\theta = d~\Delta /T_0$.

An overshooting convection configuration is built 
by imposing a two-layer piecewise-continuous polytropic
stratification (in the absence of convection) within the Cartesian domain
such that the upper layer (${0 \le
\tilde z \le d}$, layer 1) is convectively unstable and the lower layer (${d \le
\tilde z \le z_{\rm m}d}$, layer 2) is stable.  The relative stability of the
two domains is measured by the stiffness parameter, $S$
\citep{hurlburt1994penetration}, defined by
\begin{equation}
S = {m_2-m_{\rm ad} \over m_{\rm ad}-m_1},
\end{equation}
\noindent
where $m_i$ is the polytropic index of layer $i$, $m_{\rm ad} =(\gamma-1)^{-1}$ is the
polytropic index of an adiabatic polytrope, noting that $\gamma = c_p/c_v$ is
the ratio of the specific heats.  Since the polytropic index is related
to the hydrostatic heat flux and the total flux must be constant
throughout the domain, $S$ also defines a relationship between the thermal
conductivities, $\kap_i$, in the two layers, 
\begin{equation}
\frac{\kap_2}{\kap_1} = \frac{S (m_{\rm ad}-m_1) +m_{\rm ad}+1} 
{m_1+1}.
\end{equation}
\noindent
For chosen $S, ~m_1$ and $\gamma$, this conductivity contrast between the convective and stable regions is  
imposed in the simulations via a piecewise constant thermal conductivity profile as a function of $\tilde z$, with the
junction at $\tilde z=d$ smoothed by a narrow ($< 0.1d$) hyperbolic tangent function.

\subsection{Equations}
\label{sec:equations}

The equations relevant for this problem are those related to the conservation of mass, momentum and
energy, the equation of state for a perfect gas, and the induction
equation and divergence-free condition for magnetic fields.  These equations can be
non-dimensionalized using $d$ (the depth of layer 1) 
as the unit of length, the isothermal sound crossing time at the top of the domain 
$[d^2/(c_p - c_v) T_0]^{1/2}$ as the unit of time, and $T_0$ and $\rho_0$
as the units of temperature and density, to produce the full non-dimensional equation set for the problem at hand as follows:

\if\manuscript y 

\begin{eqnarray}
\label{eqn_navierstokes}
\partial_t \rho + \nabla \cdot \left(\rho \U \right) &= & 0
,\\
\partial_t \left(\rho \U \right) + \nabla \cdot \left( \rho \U \U -
\alpha_m \B \B \right)
&=&
-\nabla p_t 
-\sig C_k Ta_0^{1/2} \left(\hat{\bf \Omega} \times \rho \U \right)
\nonumber \\
& & + \sig C_k\left[ \nabla^2 \U +
\frac{1}{3} \nabla(\nabla \cdot \U) \right]
+ \rho g \hat{z}
,\\
\partial_t T + \nabla \cdot (\U T) + (\gamma - 2) T \nabla \cdot \U &=&
\frac{\gamma C_k}{\rho} \nabla
\cdot \left( {\kap_z} \nabla T \right)
\nonumber \\
& & + \frac{\zeta C_k \alpha_m (\gamma-1)}{\rho}
| \nabla \times \B | ^2  + V_{\mu}
,\\
\partial_t \B &=& \nabla \times (\U \times \B) + C_k \zeta \nabla^2 \B
,\\
\nabla \cdot \B  &= & 0
,\\
p_{t} &=& p_{g} + p_{m} = \rho T + \alpha_m \frac{|\B|^2}{2}
.\label{eqn_navierstokes_pressure}
\end{eqnarray}

\fi

\if\manuscript n 

\begin{mathletters}
\label{eqn_navierstokes}
\begin{eqnarray}
&\partial_t \rho + \nabla \cdot \left(\rho \U \right) =  0,&
\\
\nonumber \\
&\partial_t \left(\rho \U \right) + \nabla \cdot \left( \rho \U \U -
\alpha_m \B \B \right)
= -\nabla p_t &
\nonumber \\
&-\sig C_k Ta_0^{1/2} \left(\hat{\bf \Omega} \times \rho \U \right)&
\nonumber \\
& + \sig C_k\left[ \nabla^2 \U +
\frac{1}{3} \nabla(\nabla \cdot \U) \right]
+ \rho g \hat{z}, &
\\
\nonumber \\
&\partial_t T + \nabla \cdot (\U T) + (\gamma - 2) T \nabla \cdot \U =&
\nonumber \\
&\frac{\gamma C_k}{\rho} \nabla
\cdot \left( {\kap_z} \nabla T \right)&
\nonumber \\
&+ \frac{\zeta C_k \alpha_m (\gamma-1)}{\rho}
| \nabla \times \B | ^2  + V_{\mu} ,&
\\
\nonumber \\
&\partial_t \B = \nabla \times (\U \times \B) + C_k \zeta \nabla^2 \B,&
\\
\nonumber \\
&\nabla \cdot \B  =  0, &
\\
\nonumber \\
&p_{t} = p_{g} + p_{m} = \rho T + \alpha_m \frac{|\B|^2}{2}
.& \label{eqn_navierstokes_pressure}
\end{eqnarray}
\end{mathletters}

\fi

\noindent
Here, non-dimensionally, $\U = (u,v,w)$ is the velocity, $\B = (B_x,B_y,B_z)$ is the
magnetic field, $T$ is the temperature and $\rho$ is the density.
The total pressure, $p_t$, is the sum of the gas pressure, $p_g =
\rho T$, and the magnetic pressure, $p_m = \alpha_m {|\B|^2}/{2}$,
where $\alpha_m = \sig \zeta Q C_k^2$. 
The rate of viscous heating is $V_\mu =
{(\gamma -1) C_k \over \rho} \sig \partial_i u_j (\partial_i u_j +
\partial_j u_i - {2 \over 3} \nabla \cdot \U \delta_{ij} )$.

\subsection{Parameters}
\label{sec:parameters}

A set of dimensionless numbers ($\alpha_m, \sig, C_k, Ta_0,\zeta$) have appeared and other familiar ones can be derived.  These, together with the existing parameters defining the stratification $\theta, m_1, S, \gamma$ plus the aspect ratio, parameterize the problem. 

The Rayleigh number,
\label{eqn_rayleighno}
\begin{equation}
Ra(z) =
{\theta^2 (m_i+1) \over \sig C_{k_z}^2}
\left( 1 - {(m_i + 1)(\gamma-1) \over \gamma}
\right)
(1 + \theta z)^{2m_i-1} ,
\end{equation}
is a derived measure that parameterizes the competition between buoyancy driving and diffusive effects,
and thus the supercriticality and the subsequent vigor of the convection.  $Ra$ 
involves the thermal dissipation parameter $C_{k_z} = C_k \kap_z$,
where $\kap_z =  \frac{\kap_i}{\kap_1}$ and $C_k = \kap_1/\{d\rho_0 c_p
[(c_p - c_v)T_0]^{1/2}\}$; $C_{k_z}$ encapsulates the ratio of the sound 
crossing time to the thermal relaxation time in each layer. 
In this paper, $Ra(z=0.5)$ (the $Ra$ as evaluated at the middle of the unstable layer in the initial polytrope) is quoted in order to label cases.

The Prandtl number, 
\label{eqn_prandtlno}
\begin{equation}
\sig = \mu c_p/\kap_1,
\end{equation}
defines the ratio of the diffusivities of momentum and heat, evaluated in the
upper layer.  The true depth-dependent Prandtl number $\sig_z = \mu
c_p/\kap_z$ takes different values in the different layers, but note that the
diffusivity of momentum, $C_{k_z} \sig_z = C_k \sig$, 
is independent of $\kap_z$ and is therefore also independent of depth.

Next, 
\label{eqn_zeta}
\begin{equation}
\zeta = \eta c_p/\kap_1, 
\end{equation}
where $\eta$
is the magnetic resistivity, defines the ratio of the diffusivities of 
magnetic field and heat, evaluated in the upper layer.  
As for momentum, the diffusivity of 
magnetic field, $C_k \zeta$, is also independent of depth.  This parameter is often not given a specific name, and the magnetic Prandtl number, $Pm$, is a more often-quoted parameter, where here
\label{eqn_magprandtlno}
\begin{equation}
{Pm} = {Pr}/\zeta. 
\end{equation}

The Chandrasekhar number,
\label{eqn_chandra}
\begin{equation}
Q = \frac{B_0^2 d^2}{\mu_0 \mu \eta},
\end{equation}
where $\mu_0$ is the magnetic permeability, is a typical measure that quantifies the strength
of the Lorentz force of the imposed initial magnetic field $B_0$ compared to diffusive effects.     
The parameter governing the
magnetic influence in the non-dimensional equations is actually $\alpha_m = \sig \zeta Q C_k^2$ and this parameter determines the initial ratio of the gas and magnetic pressures within any imposed magnetic structure, often referred to as ``the plasma $\beta$", via $\beta =
2/\alpha_m$.  The larger $\alpha_m$, the stronger the magnetic influence.

The Taylor number,
\label{eqn_taylor_number}
\begin{equation}
Ta_0 = \frac{4 \Omega^2_0 d^4}{(\mu/\rho_0)^2}
= \left(\frac{\rho}{\rho_0}\right)^2 Ta,
\end{equation}
measures the strength of rotational effects compared to diffusive
effects.
In this paper, we generally quote $Ta$ (the more usual Taylor number) as
evaluated at the middle of the unstable layer in the initial polytrope,
for consistency with $Ra$.
In our local model setup, rotation enters the momentum equation through a modified
$f$-plane formulation via the full rotation vector,
\label{eqn_fplane}
\begin{equation}
{\bf \Omega} = \Omega_0{\hat{\bf \Omega}} = \left( \Omega_x, \Omega_y, \Omega_z \right)
= \left( 0, \Omega_0 ~{\rm cos} \phi, -\Omega_0 ~{\rm sin} \phi \right),
\end{equation}
where $\phi$ is the latitudinal positioning of the planar domain on the
sphere.
Note that with the formulation above in the $z$-downward coordinate system, positive $\Omega_0$ plus positive $\phi$ gives the
intuitive solar or terrestrial counter-clockwise rotation when viewed from above the north pole and that ${\bf \Omega}$ points north.  This is a regular right-handed coordinate system where, perhaps counter-intuitively, positive $\phi$ points in the negative $z$ direction in the northern hemisphere.  In the southern hemisphere (negative $\phi$), the coordinate system and rotation vector are aligned more as expected (${\bf \Omega}$ in the north direction aligns with the downward $z$ axis).  In this paper we will examine only the
simple polar cases where $\phi=\pm\pi/2$, representing the northern and southern hemispheres cases respectively.

{The convective Rossby number is an essential derived measure of the influence of rotation on the convective motions,
\label{eqn_rossbyno}
\begin{equation}
Ro = \left( {{Ra} \over {Ta \sig}} \right)^{1/2}.
\end{equation}
A value of $Ro < 1$ suggests a significant influence of the global rotation on the motions.}  

\subsection{Boundary conditions}
\label{sec:bcs}

The domain is periodic in all variables in the two horizontal directions, $x,y$.
At the upper and lower boundaries, it is required that
\begin{mathletters}
\label{eqn_boundaryconditions}
\begin{eqnarray}
&\rho w = \partial_z u = \partial_z v = 0\  {\rm at}\ z = 0,z_m,
\\
&T = 1 \ {\rm at}\ z=0,\ \partial_z T= (\kap_2/\kap_1)\theta \ {\rm at}\  z=z_m.
\end{eqnarray}
\end{mathletters}

\noindent
{Under these conditions, the total mass in the computational domain is
conserved and the imposed heat flux at the lower boundary is the energy source to drive the system.}  Note that the parameter $\theta$ represents the heat flux in the hydrostatic state at the bottom of the convection zone not at the bottom of the full domain.  This allows for easier comparison with purely convective simulations. 

The magnetic boundary conditions are specified
as 
\begin{equation}
\label{mag_bc_2}
\partial_z B_x = \partial_z B_y = 0\  {\rm at}\ z = 0,z_m.
\end{equation}
These boundary conditions are only imposed on the horizontal
components of the magnetic field and are sufficient, thanks to
the solenoidality of $\B$.  
The total amount
of horizontal flux within the computational domain is conserved by these boundary conditions, and they are often referred to as ``no flux'' boundary conditions \citep[e.g.][]{tobias2001transport}. 
In this paper, we are interested in the evolution of an isolated magnetic flux tube in the interior of the domain, and therefore, as long as the analysis of any simulation data is ended before the flux tube buoyantly rises to the upper boundary, these ``no flux'' boundary conditions have little influence on the outcomes of the simulations.

\subsection{Initial Conditions}\label{subsec:ics}

Magnetohydrodynamic runs are initiated using the hydrodynamic variables from a purely hydrodynamic two-layer simulation that has converged to a stationary state of turbulent overshooting convection.  Such a hydrodynamic simulation starts from the initial piecewise-polytropic state (Equations \ref{eqn_polytrope}) at the appropriately chosen parameters with small amplitude thermal noise added. 

The magnetic initial conditions consist of a pre-formed, idealized, isolated, compact, cylindrical magnetic flux structure, or ``flux tube", placed at a chosen location with a specified orientation within the domain.  We do not account for the formation of such a structure and merely impose its initial existence.  The magnetic flux tube described below will be placed within the overshoot region of the two-layer model, with the intention of mimicking the anticipated existence of such a structure created by instabilities of the larger scale dynamo magnetic field in the tachocline in the motivating solar context. 
The magnetic field $\mathbf{B}_i=(B_x,B_y,B_z)$ in the initial magnetic flux tube takes the form 

\begin{subequations}
\label{eqn_magic}
\begin{align}
    B_x(r) &= 2qB_y(r)\frac{z-z_c}{r_o}, \\
    B_y(r) &= 1 - \frac{r^2}{r_o^2}, 
        \\
    B_z(r) &= -2qB_y(r)\frac{x-x_c}{r_o}.
\end{align}
\end{subequations}

\noindent 
for $r\leq r_o$, where  $r=\sqrt{(x-x_c)^2+(z-z_c)^2}$ is the radius as measured from the center of the tube $(x_c,z_c)$ in the $x-z$ plane, and $r_o$ defines the outer edge of the finite size, isolated tube. The tube is initially a horizontally aligned cylinder, whose axis lies in the $y$ direction, and whose magnetic field is independent of that axial ($y$) direction.  The parameter $q$ is a measure of the strength of the initial twist and the sign of $q$ gives the handedness.  
The pitch angle, $\phi_p$, of a field line \citep[see for instance,][]{hughes1998rise}, relates the magnitudes of the nonaxial (locally azimuthal) and axial field within the tube ($r\leq r_o$):
\begin{equation}
    \tan\phi_p = \frac{B_{nonaxial}}{B_{axial}} = \sqrt{\frac{B_x^2+B_z^2}{B_y^2}} = \frac{2qr}{r_o}.
    \label{eqn_pitchangle}
\end{equation}

\noindent  
A twist value of $q=0.5$ therefore represents a $45^\circ$ pitch angle at the outer edge of the tube.  Positive $q$ represents clockwise twist if looking in the positive $y$ direction, and therefore positive helicity in the tube, since $B_y$ is everywhere positive.

Adding this magnetic initial condition to the hydrodynamic state requires an adjustment to the background thermodynamics in order to maintain total pressure equilibrium and to avoid acoustic transients.  Doing so provides the magnetic structure with an initial magnetic buoyancy force that initiates its rise from the overshoot layer into the convection zone above.  
That is, the magnetic field in the structure contributes magnetic pressure to
the total pressure, as seen from Equation (\ref{eqn_navierstokes_pressure}), and thus, if total pressure equilibrium is to be maintained, 
{the local gas pressure must be reduced by the equivalent amount. This may be achieved by
decreasing the temperature, density (or both) within the structure.  Here, we simply choose to decrease the local
density and to keep the temperature continuous,} assuming relatively fast equilibration of the temperature as well as the pressure, as might be expected in the low $\sig$ astrophysical environment. The fluid
inside the magnetic structure is thus less dense than its
surroundings and therefore buoyant. 
Here we examine the initially buoyant evolution of the magnetic structure in the presence of turbulent
overshooting compressible convection.

\begin{figure}[ht!]
    \centering
    \includegraphics[width=0.5\textwidth]{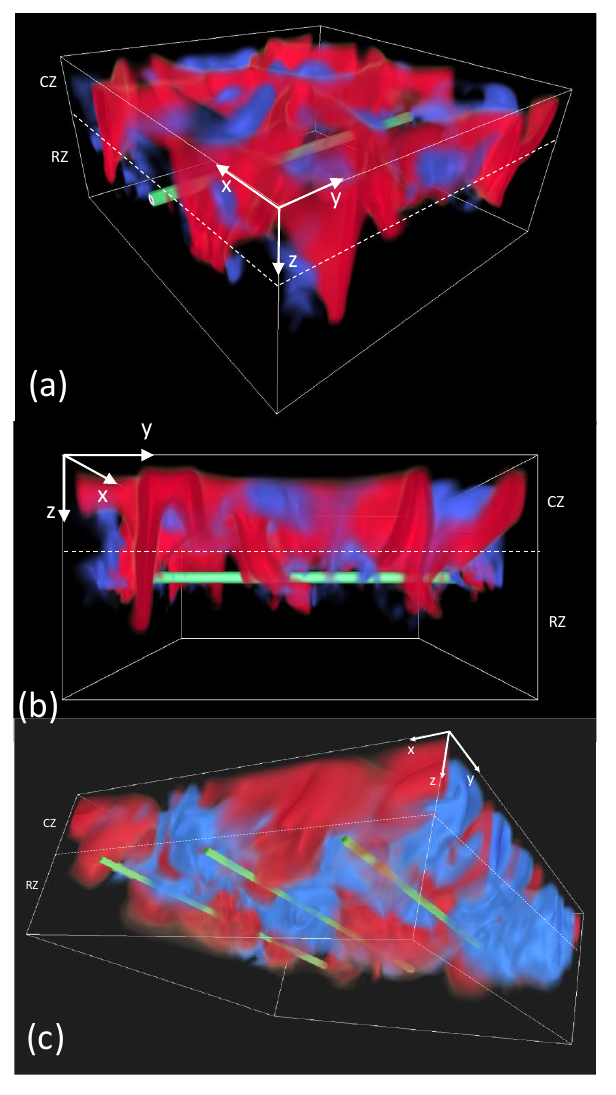}
    \caption{Volume renderings of the typical initial conditions used in these simulations. (a) An initial magnetic flux tube, with its magnetic intensity rendered in green, is placed at $x_c=3.0$, $z_c=1.25$ in a statistically stationary realization of overshooting convection, where a convection zone (CZ) overlays a radiative (or convectively-stable) zone (RZ).  The convective vertical velocities, $w$, are rendered with downflows in red and upflows in blue. The side view (b) shows that the magnetic flux tube initially sits clearly in region of overshooting convection just below the CZ. (c)  An example of a three tube magnetic initial condition. Tubes are placed at the same height as previously, but equally spaced in the periodic $x$ direction.}
    \label{fig:ic}
\end{figure}

Typical initial conditions as described above can be seen in Figure \ref{fig:ic} where we display canonical hydrodynamic turbulent convective initial conditions via the vertical velocity, $w$, with the added initial magnetic flux tubes shown in magnetic intensity, ${\bf B}^2$.   In these volume renderings, the vertical velocity is shown with red opaque tones denoting strong downflows and blue opaque tones showing the stronger upflows, and  weaker flows are made more translucent to reveal the key strong features.  Likewise, strong magnetic intensity is opaque and yellow, with green tones for weaker field.    As is typical in fairly strongly stratified compressible convection \citep[e.g.][]{cattaneo1991}, near the upper boundary the downflows form a connected network, with smooth diverging upflows between them.  The interstices of this network form strong downflows that generally traverse the whole convection zone.  In the deeper convection zone, the bulk of the motions are smaller scale and turbulent, but pierced by the strong coherent downflow plumes.  It can be clearly seen in the volume renderings that these plumes overshoot from the convection zone into the stable region below, traveling a distance that is related to the relative stiffness of the two zones, $S$, and the other parameters dictating the convective efficiency  \citep[see e.g.][]{brummell2002penetration}.  Magnetic flux tubes can also clearly be seen positioned initially in the overshoot zone, just below the majority of the convecting flows.

Figure \ref{fig:ic}$(c)$ shows an example initial condition where three tubes are initiated in a single simulation instead of just one.
Since the $\Sigma$-effect is a net statistical result emerging from turbulence, our simulation set should really constitute a Monte Carlo suite, over which we can collect meaningful statistics.   This is very numerically expensive for these complex nonlinear simulations.  In order to expedite the arrival at somewhat reasonable statistics averaged over a significant number of rising tubes, we begin many of our simulations with three identical magnetic flux tubes instead of just one as previously shown. Each of the three is oriented in the $y$-direction and seated at the same height, and are  equally spaced apart in the horizontal $x$-direction in the periodic domain.  This provides more statistics, but runs the risk of interference between the tubes during the rise.  Given our choices of parameters and scales (outlined below), we have not found this to be a problem. 

\begin{figure}[ht!]
    \centering
    \includegraphics[width=0.5\textwidth]{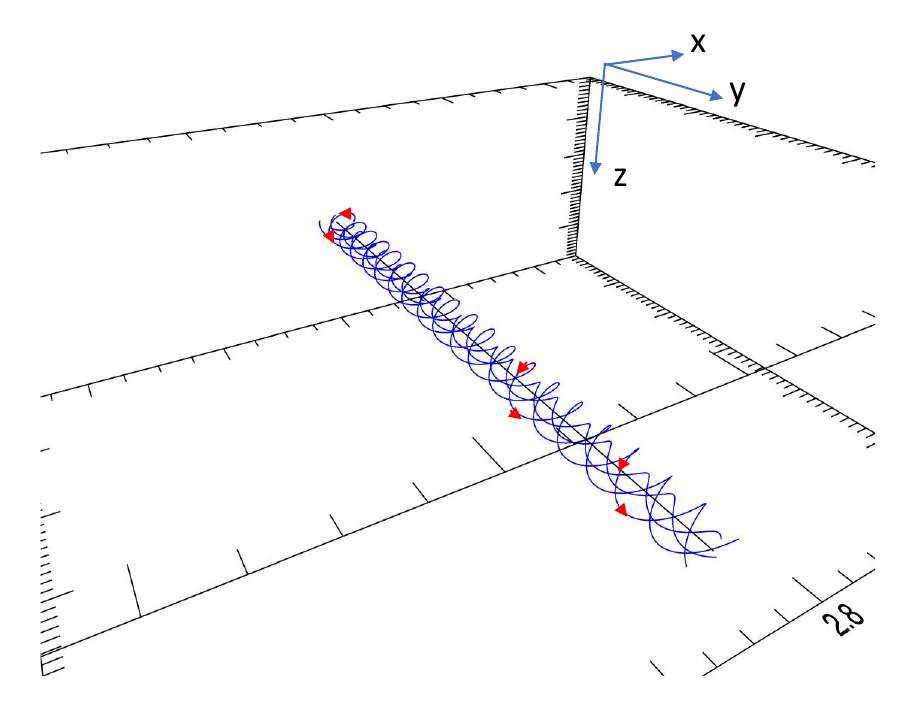}
    \caption{Several magnetic field lines (blue) of the magnetic initial condition, zoomed in on the tube within the computational domain. The magnetic field lines exhibit right-handed helicity (bearing in mind that the axial field is oriented in the positive axial $y$ direction) exhibited by the red arrows, and all loop around the central axis of the tube (black straight line in the tube) the same number of times regardless of their position within the tube, rendering the winding number of this initial condition well-defined. }
    \label{fig:ic_lines}
\end{figure}

In the accompanying Figure \ref{fig:ic_lines},  several magnetic field lines from the magnetic initial condition for a single magnetic flux tube are plotted to illustrate the topology of the field within the tube. The winding of the selected fieldlines (blue) around the central axis of the tube 
(black) leads to a net chirality indicated by the red arrows that is right-handed here for the positive $q=0.5$ chosen for this illustration and all the simulations performed in this study.


\subsection{Numerical approach}

{Equations (\ref{eqn_navierstokes})-(\ref{eqn_navierstokes_pressure}) are solved numerically as an initial value problem by a semi-implicit, hybrid
finite-diff\-erence/pseudo\-spectral scheme, using techniques and parallelized code that have been described and employed many times before
\cite[see e.g.][]{brummell1998turbulent, tobias2001transport}.  Note that these are Direct Numerical Simulations (DNS) including all diffusive processes.}

\section{Preliminaries}\label{sec:prelims}

\subsection{Basic parameters}

Our aim is to examine the evolution of magnetic flux tubes rising from an overshoot zone in the underlying convectively-stable interior as they then subsequently traverse an overlying convection zone.  We wish to assess the influence of rotation on the rise and, in particular, any relationship between the kinetic helicity of the convective state and the magnetic helicity of the magnetic flux tube as it rises.

All of our simulations presented here are performed on a $512^2\times 600$ numerical grid covering physical (dimensionless) extents $0\leq x\leq 6$ and $0\leq y\leq 6$ in the horizontal and $0\leq z\leq 2.5$ in the vertical (noting again that the convection zone is one unit deep). 
The domain is assumed to be filled with a monatomic ideal gas with $\gamma=5/3$, giving $m_{\rm ad}=1.5$.  We fix the stratification using $\theta=10$, choose $m_1=1$ so that the upper layer is superadiabatic, and assign $S=3$, corresponding to a subadiabatic lower layer with $m_2=3$.
The stiffness parameter $S$ has a strong influence on how far into the stable region convective motions from the top layer are able to reach. For large $S$, very little motion can overshoot from the upper convective layer into the lower stable layer. At low values of $S$, convective plumes from the top layer can overshoot into the bottom layer due to their inertia, creating a region of convective overshoot. The extent of this region is known as the overshoot depth, $\Delta_o$, \citep[see e.g.][]{brummell2002penetration}. Having chosen $S=3$ (and the other parameters described above and below), our simulations have an overshooting region of $\Delta_o\approx 0.5$.

We fix $Pr=0.1$ and $\zeta=0.01$ (which together imply $Pm=10$) and vary $Ra$ by altering $C_k$. We typically use $Ra=4 \times 10^4$ but also use $Ra=2 \times 10^5$ for one simulation set.  We vary $Ro$, choosing desirable values by setting $Ta$, given $Ra$ and $Pr$. We note that the critical Rayleigh number ($Ra_c$) at which convection occurs is dependent on the Taylor number ($Ta$)   \citep[for example, $Ra_c \sim  Ta^{2/3}$ for Boussinesq convection;][]{chandrasekhar1961} and so  therefore the convective vigor may also be altered. However, in reality the differences in the supercriticality of the convection due to changes in the rotation across our range of simulations are small 
so we do not compensate for this.

Our aim is to be as close to the astrophysical regime of small $Pr$, small $Pm$ and high $Ra$ as numerical issues allow, although we will be unfortunately still many orders of magnitude from true solar values.  Notably, $Pm>1$ rather than $Pm<1$ as desired.  This is because we need to reduce the magnetic diffusivity in order to keep magnetic structures from diffusing away too quickly, but then it is numerically very difficult to cope with a viscosity reduced even further.  We use $Ro$ as our major parameter for testing the $\Sigma$-effect theory, varying the influence of rotation on convection and looking for different behavior in the rise characteristics of the flux tubes.  

Our initially horizontal cylindrical twisted tubes start their evolution at depth $z_c=1.25$, which lies below the convection zone in the overshoot zone.  They start at horizontal position $x_c=3.0$ when we use a single tube, or equally spaced in the periodic domain at $x_c=(1.0,3.0,5.0)$ when we initiate the simulation with three tubes to expedite the collection of statistics.
We consistently use $r_o=0.1$ as the maximum radius of all flux tubes, and twist $q=0.5$, corresponding to sufficient twist for a tube to remain coherent during rise \citep[see e.g.][]{emonet1998physics} but not enough twist to induce a kink instability \citep[see e.g.][]{Fan98}.  The strength of the tube is given by $Q$ and the related $\alpha_m$ and we typically use 
$Q=4 \times 10^7$, leading to $\alpha_m=240$ for $Ra=4 \times 10^4$ and $\alpha_m=48$ for $Ra=2 \times 10^5$.  These are clearly all fairly low plasma $\beta$ ($\sim \alpha_m^{-1}$) calculations.  Typically, the deep interior of the Sun is thought of as a high $\beta$ environment due to the significant stratification and therefore high gas pressure at depths.  However, almost by definition, the local $\beta$ must be low in a compact structure for it to rise by magnetic buoyancy. Note also that $Q$ cannot be set arbitrarily high in these types of simulations since this might try to force very low or unphysical negative densities during the thermodynamic adjustment to the insertion of the magnetic structure.  

Note that some of the important parameters of this problem are difficult to choose since we know so little about the deeper solar interior.  
Even the initiation location of rising magnetic structures is unknown, although prevailing wisdom seats the origin in the tachocline \citep[e.g.][]{strugarek23}, and it is likely that it is at least deep in the convection zone \citep[e.g.][]{Nelson2014}.  We place our structures initially in the overshoot zone of the convection.

In particular, we do not know the relative scales of the velocity and magnetic fields.  Our model convection zone has a density contrast across it of $\chi_\rho=\rho(z=1)/\rho(z=0)=11$ initially (but relaxes to $\chi_\rho \sim 20$ as the simulation evolves), and covers $\sim 5$ pressure scale heights, thereby corresponding to perhaps the lower half of the solar convection zone.  We are certainly not simulating solar convection near the surface, where the density drops very rapidly, only instead following deep interior dynamics.  The convective scales in our model are therefore not representative of solar granulation nor supergranulation, but rather deeper and probably much larger overall scales.  This is important for our choice of rotational influence, since we desire $Ro < 1$ in order to experience any significant rotational effects on the convection.  For the Sun, $Ro \sim 1$ for so-called giant cell solar convection scales of order the depth of the convection zone, but $Ro \sim 10$ for supergranular scales of about $\sim 30$Mm and $Ro$ is much higher for granulation at scale $\sim 1$Mm. 

Furthermore, our observational information on emerging magnetic structures and their scales is limited to the near-surface of the sun.  Sunspots vary greatly in size but are generally much larger than granulation, more typically on the order of supergranulation. Active regions can expand up to significantly larger sizes.  We might expect the diameters of sunspot-generating magnetic structures to be significantly smaller in the deeper interior, thanks to the stratification.  Here, we have therefore made the radius of our initial magnetic flux structure substantially smaller than a typical convective cell width (measured at the model upper boundary), but still a significant fraction.  Our tube radius is $\sim 1/10$ of a surface cell size (although this ratio varies with some of the parameters, such as $Ra$ and $Ro$).  

Similarly, there is no good observational information about the relative strength of the magnetic field at the initiation site for the rising magnetic structures.    It is expected that the plasma $\beta$ must be low in rising compact structures, as mentioned earlier, and from simulations \citep[e.g.][]{Cline:thesis:2003, Abbett2004} it is known that the magnetic energy in a flux tube must be significantly higher than the peak kinetic energy in the downflows of the convection for a flux tube to survive via buoyant rise.  Using these guidelines and our own experimentation, we choose our initial field tube strength to guarantee survival during transit and yet provide a relatively slow buoyant rise rate.  The actual flux tube strength required for slow, guaranteed rise depends on the other parameters, such as $Ra$ and $Ro$, but we try and give the $\Sigma$-effect as much time as possible to operate, while keeping $Q$ constant for all cases for ease of comparison.  In general, our ratio of rise time to convective turnover time, $\tau = \tau_{\rm rise}/\tau_{\rm turn}$, ranges from $\tau \sim 1$ for the largest (domain) scale
to $\tau \sim 50$ on the smallest numerically-resolved, non-dealiased scale.
Assuming that most of the work of helicity transfer is perhaps done at some intermediate turbulent scales, this ratio seems reasonably high, i.e. the $\Sigma$-effect has plenty of time to operate.  It is worth noting here that, in the highest $Ra$ cases performed at  $Ra=2\times10^6$, the flux tube rise times are much longer and therefore test the extension of this ratio.  
{Overall, in all our simulations, we chose $Q$ such that our tubes are superequipartition to the strongest flows. irrespective of all other parameters (such as $Ra$) that affect the kinetic energy.}

\subsection{Measures}

The underlying premise of the $\Sigma$-effect is that the twist of a magnetic flux structure is indirectly affected by the convection, via the effect of the kinetic helicity of the turbulence on the writhe.  Our simulations necessarily start with a magnetic flux tube with a prescribed, sufficient amount of twist, otherwise a coherent rise of the tube does not occur even in the absence of convection.  We therefore need to quantify the evolution of the amount of twist within the magnetic structure during its magnetically-buoyant rise and interaction with the convection.

Despite the fact that we are using fully resolved 3D simulations of the dynamics and therefore have access to the full magnetic field, measuring the helical properties of a moving magnetic structure can still be quite tricky. We attempt to measure the degree of twist using a winding number \citep[see, for instance,][]{prior2014helicity, berger2006writhe}, defined as follows. If we consider two distinct magnetic field lines, $\mathbf{f}_1$ and $\mathbf{f}_2$, that initially wind in the positive $y$ direction in our initial condition, and assume that the convection contorts them sufficiently little that they still always travel in the positive $y$-direction (i.e. do not fold back on themselves), then the curves can be parameterized uniquely in $y$ and the vectors $\mathbf{f}_1(y)$ and $\mathbf{f}_2(y)$ contain the $(x,z)$ locations of the lines at every $y$ position, i.e. $\mathbf{f}_i(y)=(x_{f_i}(y),z_{f_i}(y))$. 
That this assumption is true can be verified after the fact (see e.g. Fig. \ref{fig:sh_lines}).
The winding number, $\mathcal{L}$, is then defined as the net rotation of the vector $\mathbf{r}=\mathbf{f}_2-\mathbf{f}_1$ as we move along the tube from $y=0$ to $y=y_m$, given by
\begin{equation}
    \mathcal{L}(\mathbf{f}_1,\mathbf{f}_2) = \frac{1}{2\pi}\int^{y_m}_{0} \frac{d}{dy}\Theta(\mathbf{f}_1(y),\mathbf{f}_2(y))dy,
    \label{eqn_wi_1}
\end{equation}

\begin{equation}
    \text{where \ }\Theta(\mathbf{f}_1,\mathbf{f}_2) = \arctan\left(\frac{z_{f_1}-z_{f_2}}{x_{f_1}-x_{f_2}}\right).
    \label{eqn_wi_2}
\end{equation}

\noindent Here then, $\Theta(\mathbf{f}_1,\mathbf{f}_2)$ measures the angle between the two lines at every point along the the $y$-direction from $0$ to $y_m$ and $\mathcal{L}$ counts how many turns that angle makes, i.e. the number of times the lines $\mathbf{f}_1$ and $\mathbf{f}_2$ wind around one another. 

If we select the line $\mathbf{f}_1$ to be a line that defines the center of the flux tube, $\mathbf{f}_c$  (noting that it is not necessarily a field line), and $\mathbf{f}_2$ to be any other fieldline in the tube, then $\mathcal{L}$ represents the winding of that fieldline about the tube center, and thus is a measure of the twist of this particular field line within the flux tube.   We calculate this measure for a (large) number of field lines within the flux tube and average to arrive at our overall measure, namely the average winding number of fieldlines within the tube around the centerline, 
\begin{equation}
    \mathcal{W}=\sum_{i=1}^{N}\mathcal{L}(\mathbf{f}_c,\mathbf{f}_i)/N,
    \label{eqn_windingnumber}
\end{equation}

\noindent where 
the vector $\mathbf{f}_i$ represents the $i$th fieldline and 
$N$ is the number of field lines over which we are averaging.  

We are generally interested in changes to the winding number measure, $\mathcal{W}$, over time.  During the evolution of simulations, we dump data for the full three-dimensional magnetic field vector, $\mathbf{B}$, at a chosen time cadence.  For each time dump, we integrate the field lines, $\mathbf{f}_i$, using a standard second-order Runge-Kutta scheme where we select seeds at $y=0$ and integrate through the $y$-direction to $y=y_m=6$. The seeds are selected on the grid for every point $(x,0,z)$ that meets the criteria that the magnetic energy is within $80\%$ of the maximum magnetic energy in that $y=0$ plane. This cutoff identifies locations that are within the flux tube, provides a substantial number of field lines for integration and therefore a reasonable average, and generally omits weak field lines that tend to have more complex topology. 

The $(x,z)$ locations of the center line, $\mathbf{f}_c$, as a function of $y$ are selected as the location of the maximum magnetic energy in each $y$-plane on the numerical grid. We found that, in our simulations, this choice consistently produces well-defined locations and creates a smooth line to represent the center of the tube.  As we will discuss in more detail later, when the flux tube interacts with convection, its originally axially ($y$) invariant shape will become distorted (writhed), and defining the center of the tube can become more ambiguous but our procedure seems to hold up in general. 
{We note that calculating a value for the net writhe of the magnetic structure (the winding of the central axis around itself) is much more complex and we have not attempted to do this, although we do show the axis and its writhe (as projections onto the coordinate plane walls) on the fieldline plots that follow.}

This method of calculating a winding number can at least be verified against the known analytical initial conditions. The equation for the pitch angle of the fieldlines within the tube in the initial conditions is given in Equation (\ref{eqn_pitchangle}).
The axial ($y$) distance, $\ell$, that a field line traverses for a full rotation around the tube is thus $\ell=\pi r_o/q$, and therefore the number of times a fieldline will wind around the tube when traversing the $y$-direction from $0$ to $y_m$ is $\mathcal{W}_{ic}=y_m/\ell$.
Notice that this is independent of $r$, i.e. our initial conditions were chosen to have constant winding number for all fieldlines.  For our chosen canonical parameters $y_m=6, r_o = 0.1, q=0.5$, this results in $\mathcal{W}_{ic}\approx 9.55$. Using our numerical method leading to Equation (\ref{eqn_windingnumber}) above on the numerical data for the initial condition reproduces results within $\pm 0.01$ of this value. Obviously this is just a very simple sanity check and we must be aware of the possible drawbacks that this numerical method may have for more complex magnetic field topology.  We also comment that it may at first seem counterintuitive that $\mathcal{W}_{ic}$ and $\mathcal{W}$ do not necessarily take integer values, since our lateral boundary conditions are periodic.  Note however, that since any component of the magnetic field in the flux tube has a maximum somewhere within the flux tube and drops to zero at the edges, there are closed contours of any value within that range enclosed in the tube, and therefore that component of the field can complete periodically anywhere on that contour.  Thus, any real value of the winding number is allowed, and arbitrary amounts of twist may be gained or lost by the magnetic structure.


\begin{figure}[ht!]
    \centering
    \includegraphics[width=1.0\textwidth]{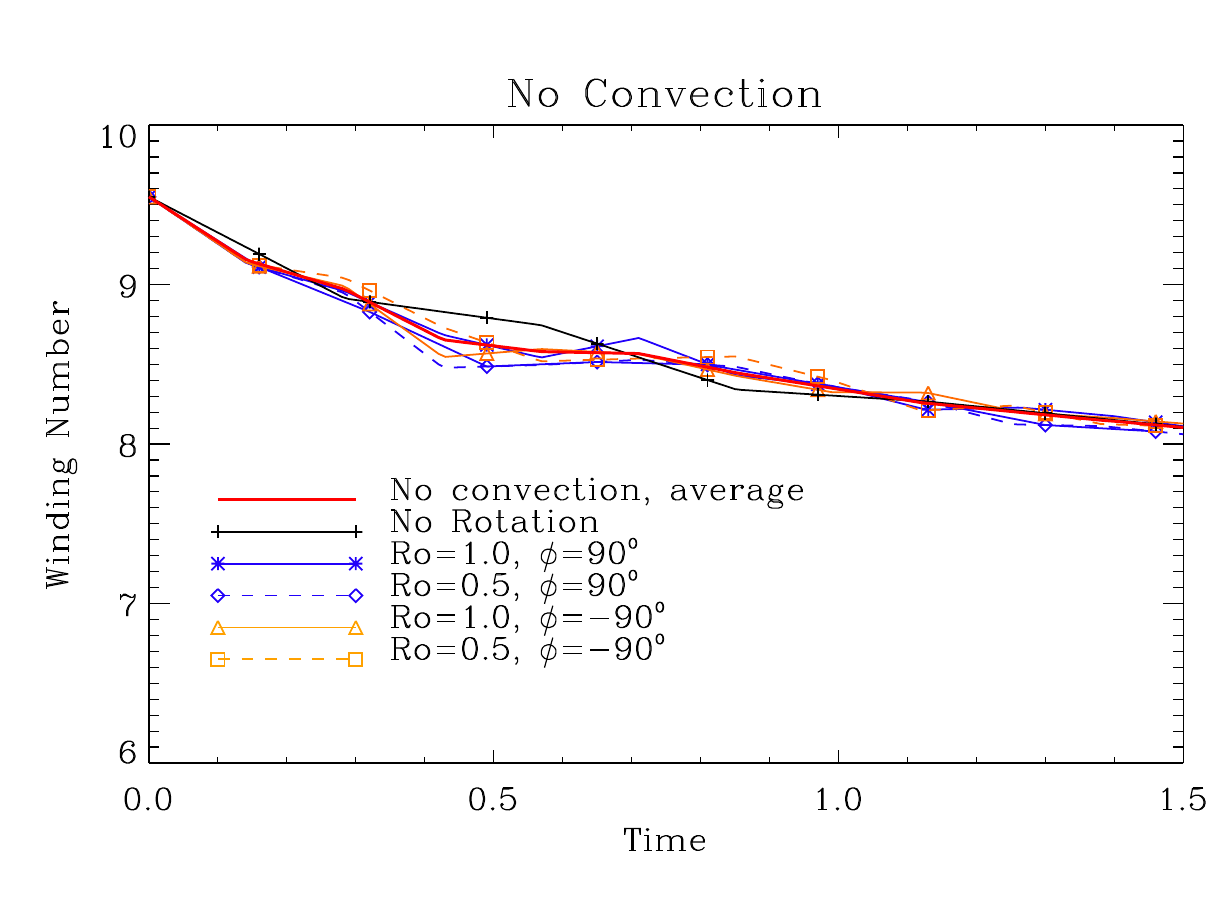}
    \caption{Diffusive decay of $\mathcal{W}$ versus time for five different parameter pairs ($Ro, \phi$) in the absence of convection. Parameters of the simulations plotted encompass $Ro=0.5,1.0$ with $\phi=+90^\circ$ and $\phi=-90^\circ$, as well as a case with no rotation ($Ro=\infty$).
    The colored lines associated with each case here will pervade throughout this paper.   The nearly identical decay of $\mathcal{W}$ at various rotational influences suggests that the decay of $\mathcal{W}$ in the absence of convection is purely diffusive and is independent of rotation. The red line shows the average non-convective decay of $\mathcal{W}$ which will be used as a reference for later plots.}
    \label{fig:W_diff}
\end{figure}

\section{Results}
\label{sec:results}

\subsection{Evolution of \texorpdfstring{$\mathcal{W}$}{W} without convection}

Although the direct numerical simulations of this study are at  moderately low Prandtl numbers and moderately high Reynolds numbers (and are at relatively high resolutions), the presence of molecular diffusion can still be important due to the presence of significant field gradients. 
Since we are dealing with a compact, isolated magnetic structure (the flux tube), where the magnetic field component amplitudes drop algebraically in radius to zero everywhere outside of the radius of the tube, the field components initially have a discontinuous derivative at the edge of the tube.  We therefore might expect the tube to diffuse significantly, at least initially, and this may have an impact on the evolution of $\mathcal{W}$. We wish to be able to distinguish such effects from the effects of rotationally-influenced convection, and so we first examine simulations in the absence of convection as a calibration. 

In order to do this, we evolve Equations (\ref{eqn_navierstokes}-\ref{eqn_navierstokes_pressure}) away from initial conditions where all velocities are set to zero but the magnetic field is still set to include the flux tube initial condition specified by Equations \eqref{eqn_magic}.   We apply our canonical parameters described above.
It may seem necessary to lower $Ra$ such that the upper layer is convectively stable but the rise time of the tube is fast compared to the onset time of the convective instability, so starting from static initial conditions with no numerical perturbation actually allows us to keep all the parameters exactly the same as for the following simulations including $Ra$. We repeat this non-convective exercise at various 
$Ro$ and $\phi$ for peace of mind, even though we do not expect these to have any significant effect.  

Figure \ref{fig:W_diff} displays the values of $\mathcal{W}$ versus time as lines with both color and symbols for a set of simulations with $Ro=(1.0,0.5)$ at both $\phi=\pm90^\circ$ (corresponding to both northern and southern hemispheres) and a case with no rotation ($Ro=\infty$).  Note that both the color and symbols associated with the parameter pair values ($Ro, \phi$) will be maintained in all figures throughout this paper for clarity, accessibility and continuity. As expected, the resulting decay of $\mathcal{W}$ in each case here overlap very closely, likely differing mainly within the range of the numerical errors incurred during the calculation of the fieldlines, $\mathbf{f}_i$. The average of all of these simulations is shown by the solid red line.  This confirms the notion that $\mathcal{W}$ will evolve somewhat due to diffusive effects in the absence of convection, but also that the decay appears to be independent of rotational effects, even though there are some differences in the magnetic buoyancy-induced flow field for the different cases.  
Any velocity vectors that are not purely vertical associated with the buoyant rise are deflected by Coriolis forces by an amount that depends on the $Ro$.  This causes tilting of the flow lines and a subsequent decrease in the vertical flow component \citep[see also][]{hughes1985magnetic, ziegler2001effect}. We therefore observe slower rise speeds at larger values of $Ta$ (lower $Ro$ at fixed $Ra$ and $Pr$) and the fastest rise speeds with no rotation. However, these changes in the rise speed of the tube do not seem to play a significant role in the resistive decay of $\mathcal{W}$ over time. 

It is important to note that the density perturbation which induces buoyant rise in the flux tubes in our simulations is invariant along the axis of the tube (independent of the $y$ direction).  In this case,  during the evolution of the non-convective simulations, the tube remains in a perfectly horizontal orientation, similar to the initial condition (but diffusively spread somewhat).  The flux tube does not become deformed along its axis in any way in these non-convective cases, indicating the lack of any undular or kinking instability. If a density perturbation that did vary in the $y$-direction were to be introduced initially, it would result in differential rise of portions of the tube, potentially creating an artificial $\Omega$-loop type structure.  Alternatively, if the twist were to be set sufficiently stronger than we have chosen, a kink instability could occur, potentially also producing an $\Omega$-loop in the tube.  Such a configuration can result in a writhe of the tube's axis simply by the action of the Coriolis force on the flows within the ``legs'' of the $\Omega$-loop. This result has been shown to be consistent with Joy's law and can actually result in the correct sign  of twist for the SHHR (\cite{longcope1997dynamics}, hereafter LK). However, LK points out that the resulting writhes in their models are too small to match the twist values of the SHHR observations, and, furthermore, that the model does not account for the statistical scatter of the SHHR. 
In our context, these potential alternative methods of writhing (undular instability modes or kinking) are undesirable. In order specifically to study purely the $\Sigma$-effect in our simulations here, we only consider flux tubes where the density perturbations are (at least initially) invariant along the axis of the tube and the twist is sufficiently low to prevent kink instability.

\subsection{General Dynamical Evolution of Convective Simulations}

We now discuss a suite of simulations where we concentrate on the evolution of the winding number, $\mathcal{W}$, in magnetic flux tubes rising buoyantly through a rotationally-influenced convection zone. The magnetic initial conditions to create the flux tube are the same as in the previous non-convective simulations (Equations \ref{eqn_magic}), but now the hydrodynamic variables are initialized from stationary states of previously-computed overshooting convection cases that are subjected to varying degrees of rotational influence.   We focus on the influence of rotation by varying $Ro$ and the hemisphere in question, choosing either $\phi=\pm 90^\circ$ to represent the northern or southern hemisphere respectively.  If the $\Sigma$-effect holds, then we expect differences in the winding numbers, $\mathcal{W}$, depending on these parameters.

We note again that for any given parameter set ($Ra$, $Ro$, $\phi$), we run a Monte Carlo suite of simulations consisting of at least 6 instances.  Each instance corresponds to either a different hydrodynamic initial condition from the time series of potential convective initial conditions of the precursor convective simulation, or a different spatial location for the magnetic flux tube in the same hydrodynamic initial condition when three tubes are tracked at once in the same simulation.


\begin{figure}[ht!]
    \centering
    \includegraphics[width=1.0\textwidth]{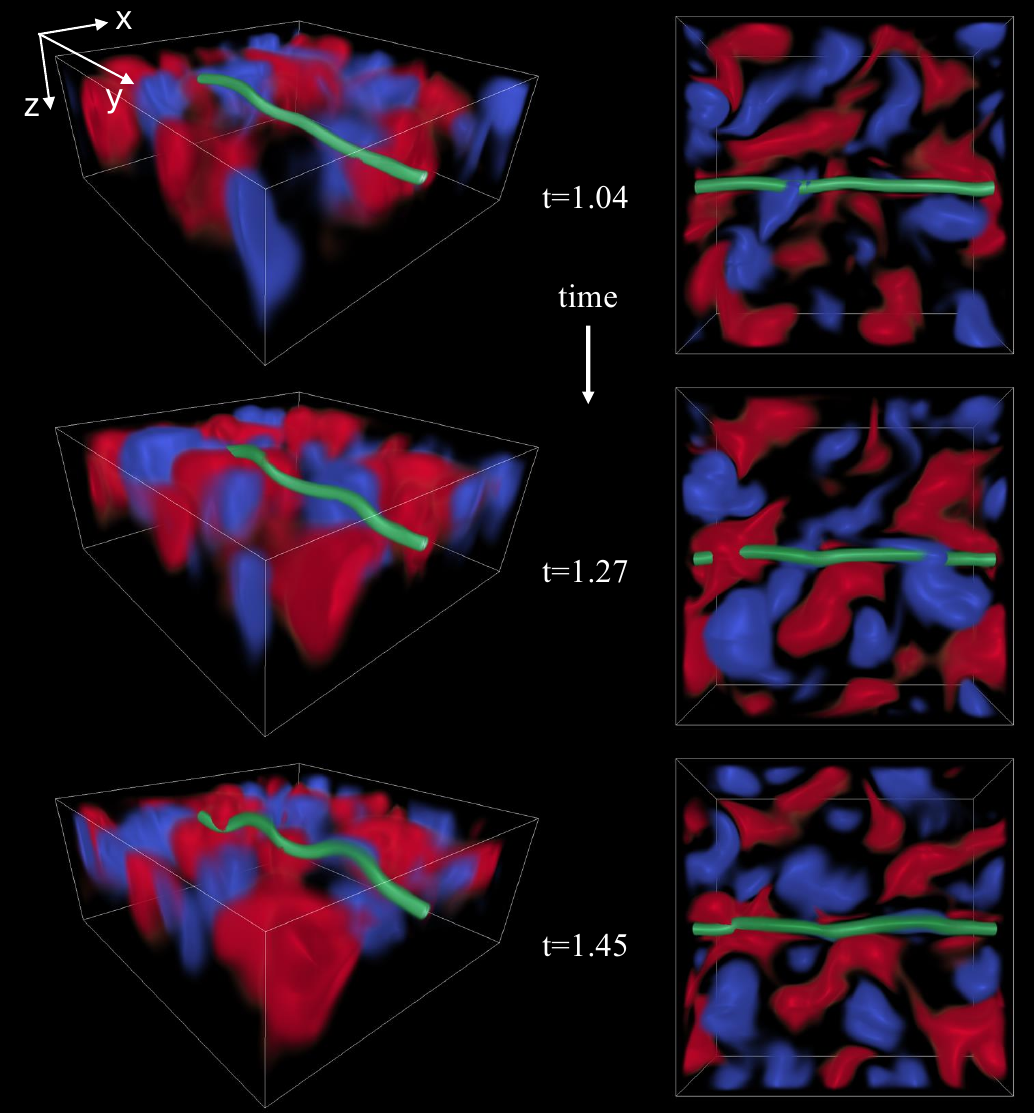}
    \caption{Example of the typical evolution of a simulation.  Volume renderings of both the magnetic intensity, $\bf{B}^2$ (green=yellow tones), and the vertical velocity, $w$ (shown with downflows in red tones and upflows in blue).  In the volume rendering, the opacity is set to emphasize high (absolute) values of each field.  Time evolves from the top row to the bottom for the two views shown (angled view and top down view).   This case is shown here simply as a clear example of a magnetic flux tube rising through rotationally-influenced convection, but is the case with $Ro=1.0$ in the southern hemisphere ($\phi=-90^{\circ}$) used later in this paper. The times shown are $t=1.04, 1.27, 1.45$ (top to bottom).  }
    \label{fig:sh}
\end{figure}


\begin{figure}[ht!]
    \centering
    \includegraphics[width=1.0\textwidth]{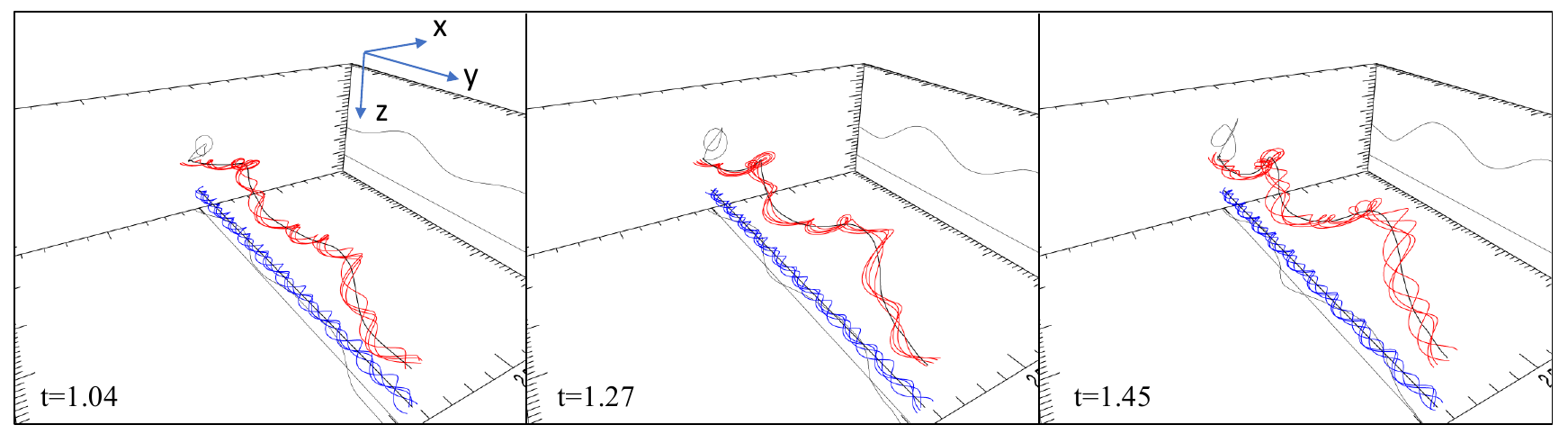}
    \caption{A further example of the general evolution of a simulation, this time shown via the magnetic fieldlines.  This case corresponds exactly to the parameters and times of the previous figure, Fig. \ref{fig:sh} ($Ro=1.0, \phi=-90^{\circ}, t=1.04, 1.27, 1.45$.  Several randomly chosen magnetic field lines from the many used in the calculation of $\mathcal{W}$ are shown in red.  Several randomly chosen field lines of the initial condition are also shown in blue for reference. The black line shows the axis of the tube, discerned as the peak magnetic intensity in the $x-z$ plane at any $y$.  The grey lines on the coordinate axis planes show the projection of the flux tube axis onto the respective planes.  These examples demonstrate the clear deformation of the magnetic flux tube as it rises through the rotationally-influenced convection.}
    \label{fig:sh_lines}
\end{figure}

Figures \ref{fig:sh} and \ref{fig:sh_lines} show examples of the general dynamics of the simulations.  These figures were simply chosen as clear examples, but for reference are at parameters $Ra=4\times 10^4, Ro=1.0$ and $\phi=-90^\circ$ (southern hemisphere).  These figures are referenced again in later sections. 
Since the initial configuration with the insertion of the magnetic structure (shown in Figure~\ref{fig:ic}) is deliberately not an equilibrium state, the magnetic structure immediately begins to rise out of the overshoot zone and through the convection by magnetic buoyancy. This is clearly shown in the angled-view volume renderings of the evolution in the left side of Figure \ref{fig:sh}. In this figure, both the vertical velocity and the magnetic intensity are rendered together. As the magnetic structures rise, they interact with the convective flows and become deformed, losing their initial $y$-direction axial invariance and becoming contorted and potentially writhed, as can also be seen in the right hand side time evolution in the figure. 

Figure \ref{fig:sh_lines} shows the equivalent time steps in the evolution of this case as magnetic field lines.  The red lines displayed in each panel show selected field lines from the many field lines used in the calculation of $\mathcal{W}$. The black line amongst the red lines is not a field line and instead marks the center of the magnetic structure as defined by the peak magnetic intensity in any $y={\rm constant}$ plane. The grey lines on the sides of the domain are the projections of this center line of the magnetic structure onto that co-ordinate axis plane. These give a clear indication of the degree of convective deformation of the flux tube from the perfectly straight ($y$ independent) initial condition. 
This example shows a typical evolution, with a slight deformation as the magnetic structure enters the convection zone near $z=1$ ($t=1.04$), but by the last time step ($t=1.45$) the structure is significantly deformed. The major observable deformations in the final frames of these cases appear to be visually correlated with the presence of significant strong convective large-scale downflows rather than any other systematic effect, as we will now attempt to quantify for various cases using the winding numbers, $\mathcal{W}$.


\begin{figure}[ht!]
    \centering
    \includegraphics[width=1.0\textwidth]{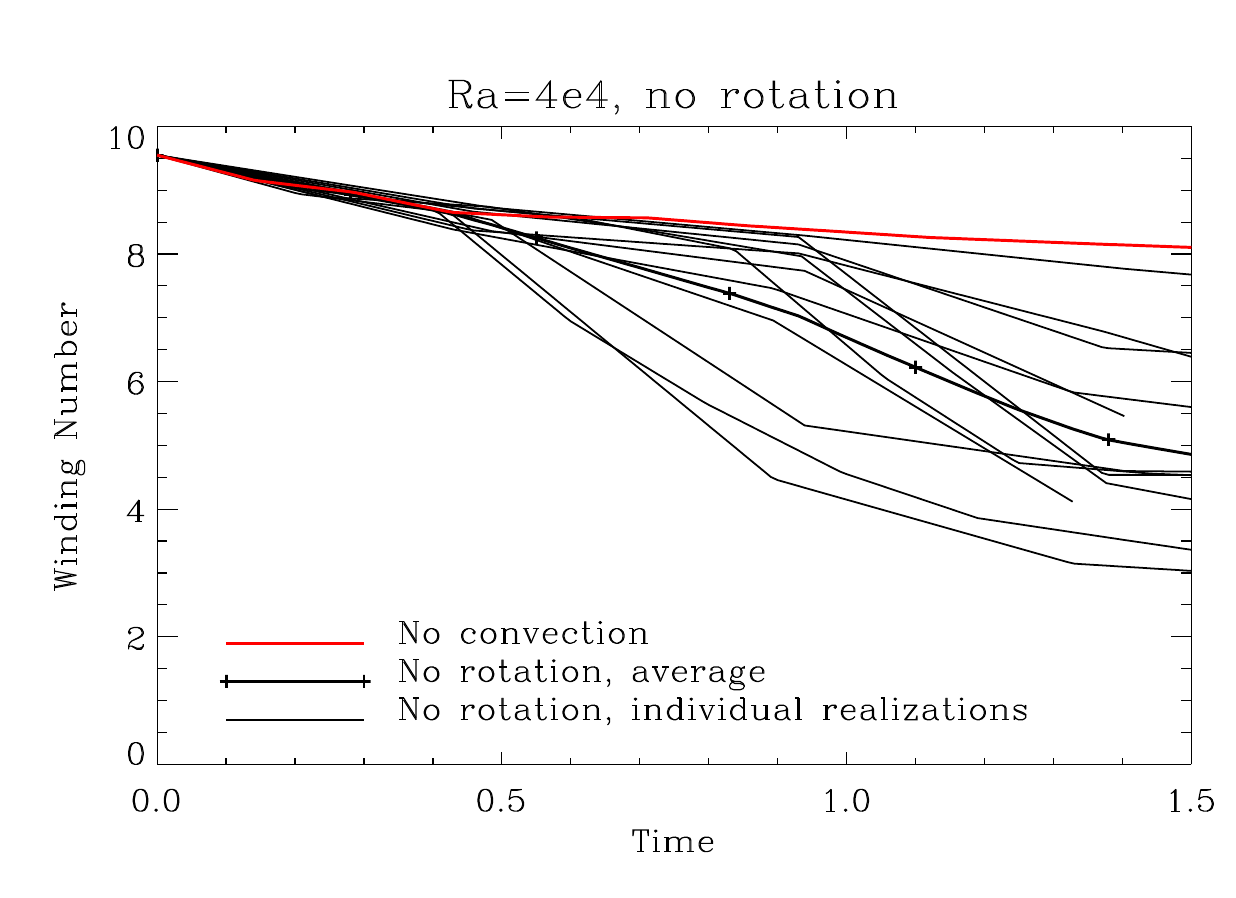}
    \caption{Evolution of $\mathcal{W}$ versus time for non-rotating convection. Each black line without a symbol corresponds to the evolution of an individual tube embedded in non-rotating convection at $Ro=\infty$ and $Ra=4\times 10^4$.  The average evolution is displayed in thicker black with + symbols. The non-convective diffusive decay is shown in red for reference.  Clearly, convection generally causes faster decay of the winding number.}
    \label{fig:W_noRot}
\end{figure}


\begin{figure}[ht!]
    \centering
    \includegraphics[width=1.0\textwidth]{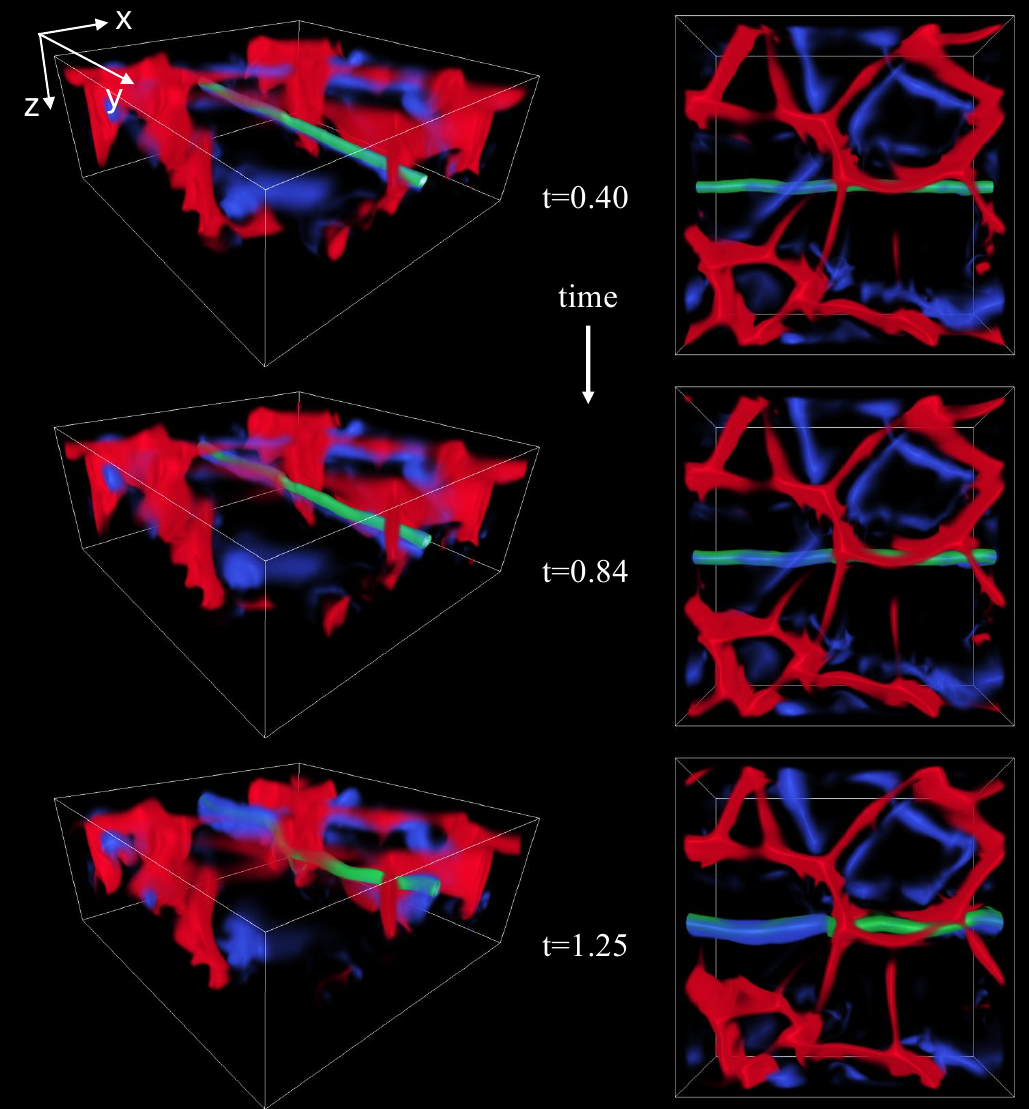}
    \caption{An example of a magnetic flux tube rising in the presence of non-rotating convection. The format is the same as the example figure, Fig. \ref{fig:sh}.  Shown are times $t=0.40, 0.84, 1.25$ (top to bottom).}
    \label{fig:noRot}
\end{figure}

\subsection{Effect of non-rotating convection on \texorpdfstring{$\mathcal{W}$}{W}}

In order to quantify the change in writhe and twist of a magnetic structure due solely to convective effects {\it in the absence of rotation}, we first plot in Figure \ref{fig:W_noRot} the resulting evolution of $\mathcal{W}$ from a significant number of simulations of the flux tube rise dynamics in non-rotating convection.  The (average) purely diffusive (non-convective) evolution of $\mathcal{W}$ from before (Fig. \ref{fig:W_diff}) is shown for reference in red. Each black line without a symbol corresponds to the evolution of $\mathcal{W}$ extracted from a simulation at the same parameters but starting from a different realization of the hydrodynamic initial conditions.  The time evolution of each realization is cut off as the magnetic structure approaches the upper boundary of the domain, and results are only shown where there is confidence that there is no influence of the upper boundary.  The thicker black line with + symbols corresponds to the average evolution of $\mathcal{W}$ over all the realizations at these parameters.  The evolution of a typical realization is shown in the volume renderings of Figure \ref{fig:noRot} in the same style as previously shown (e.g. Figs.~\ref{fig:ic} and \ref{fig:sh}).  

In Figure \ref{fig:W_noRot}, the early evolution of $\mathcal{W}$ remains relatively consistent with the purely diffusive evolution for all realizations until around $t=0.5$.  This time period corresponds to the initial rise of the magnetic structure through the calmer overshoot zone into the convection. The first row of Figure \ref{fig:noRot} confirms that the initial evolution of the tube remains mostly invariant in the $y$-direction during this period. Somewhere between $t=0.5$ and $t=1.0$ depending on realization, the value of $\mathcal{W}$ generally drops fairly significantly from around $\mathcal{W}=8.5$ to somewhere between $\mathcal{W}=3$ and $\mathcal{W}=7$ by $t=1.5$. There is significant scatter, but the average loss of $\mathcal{W}$ at $t=1.5$ compared to the non-convective, diffusion-only case average value is about 40\%.  During this time, it can be seen that there is a significant deformation of the central axis of the tube by the convection, as exhibited in the second and third rows of Figure \ref{fig:noRot}.  The tube becomes increasingly deformed as it interacts with the convection higher up in the domain. Strong convective downflows (red) can be seen to influence the deformation of the magnetic structure significantly.  For example, in the final row of Figure \ref{fig:noRot}, at the left end of the flux tube where there are no (red) downflows, the magnetic structure (shrouded in its own buoyant blue upflows) can be seen to have risen significantly more than the right end of the tube, where strong downwelling is holding down this section of the flux tube.

From the $\Sigma$-effect theory, we expect no systematic bias in the evolution of $\mathcal{W}$ in these simulations where no rotation is present.   We might have anticipated that there should be only random helical velocity perturbations and therefore ultimately equal amounts of both senses of writhe generated in the tube axis and twist increments subsequently induced, at least in some sufficient average.  We see that this is not the case here though, with a definite loss of $\mathcal{W}$, even more than the purely diffusive cases, and no detectable gain.  It is tempting to simply attribute this to non-helical vertical stretching of the flux tube in the interactions with strong convective downflows, but here we calculate the total $\mathcal{W}$ over the length of the tube and not the value per unit length, where such an effect would show up as a loss.  It seems as though the culprit is still likely interactions with the strong downflows, but that the advective interactions act more as a turbulent diffusion, disrupting the tube and enhancing diffusive processes.  Regardless of origin, it is important to use these non-rotating results to calibrate later simulations where we seek the effect of rotation.  


\begin{figure}[ht!]
   \centering
    \includegraphics[width=0.65\textwidth]{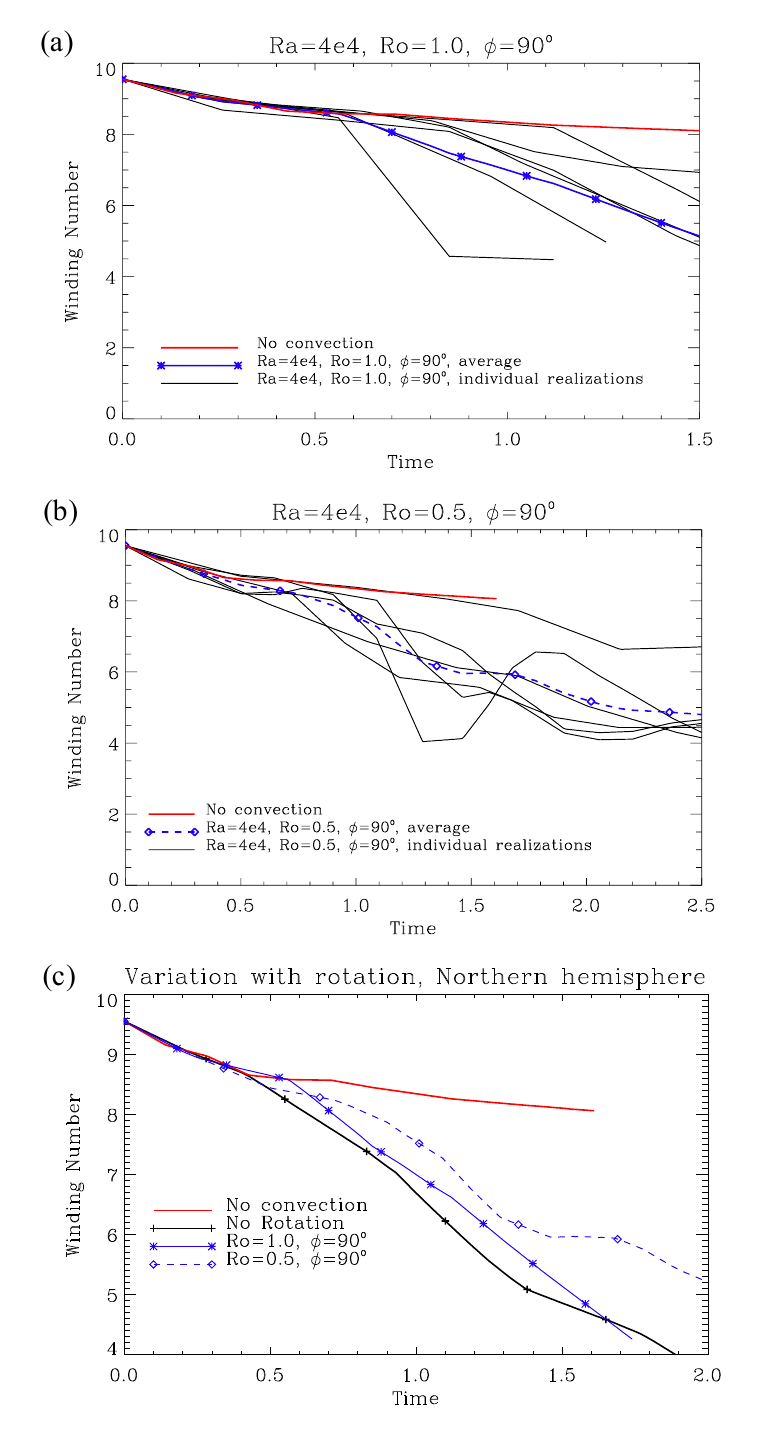}
    \caption{Evolution of $\mathcal{W}$ versus time for rotating convection at $Ra=4\times 10^4$ in the northern hemisphere ($\phi=90^{\circ}$) for various rotation rates: $(a)$ $Ro=1.0$ and $(b)$ $Ro=0.5$. Each black line without a symbol corresponds to the evolution of an individual tube. The average over all realizations in each case is displayed thicker, in color, and with a symbol: blue with a 8-spoked symbol for $Ro=1.0$ and dashed blue with diamonds for $Ro=0.5$. Panel (c) compares the average evolutions of $\mathcal{W}$ at each $Ro$ with the non-rotating case ($Ro=\infty$; black with + symbols) and the non-convective case (red).}
    \label{fig:W_north}
\end{figure}

\subsection{Effect of rotating convection on \texorpdfstring{$\mathcal{W}$}{W}}

\begin{figure}[ht!]
    \centering
    \includegraphics[width=1.0\textwidth]{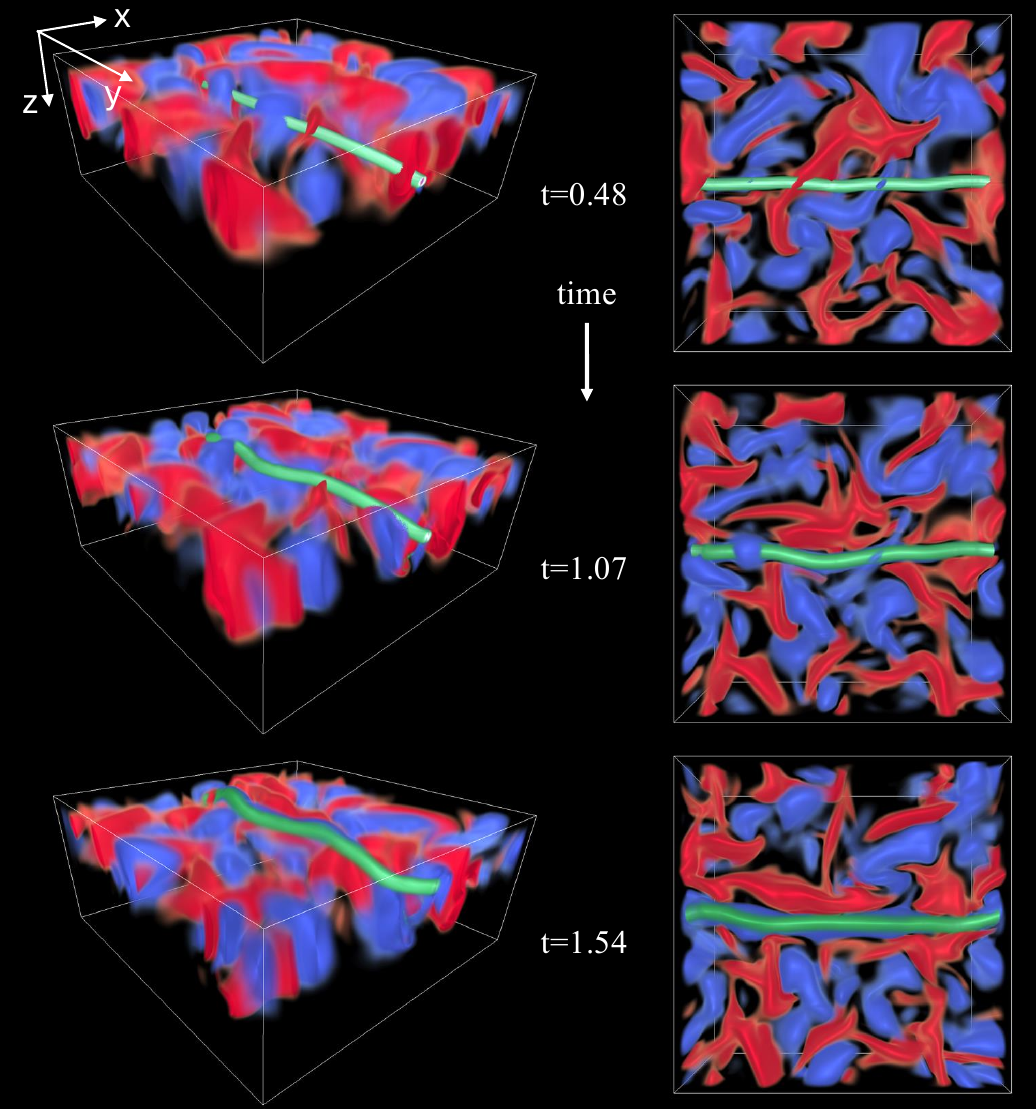}
    \caption{An example of a magnetic flux tube rising in the presence of rotating convection at $Ro=1.0$ in the northern hemisphere ($\phi=90^{\circ}$). The format is the same as the example figure, Fig. \ref{fig:sh}.  Shown are times $t=0.48, 1.07, 1.54$ (top to bottom).}
    \label{fig:nh}
\end{figure}

\begin{figure}[ht!]
    \centering
    \includegraphics[width=1.0\textwidth]{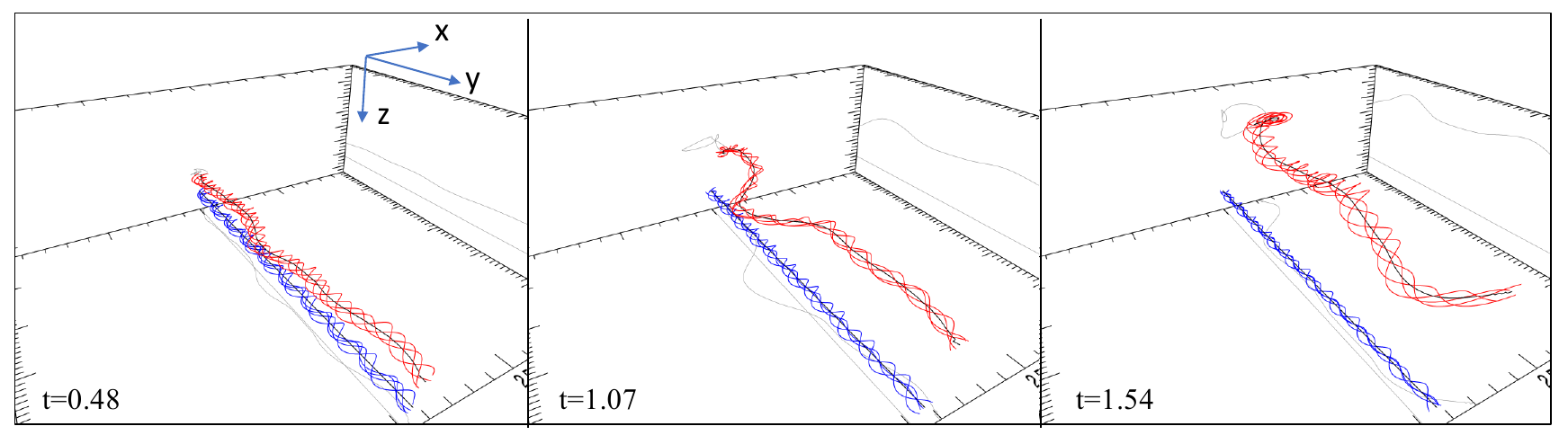}
    \caption{Illustration of the field line contortion for the case in the previous figure, Fig.\ref{fig:nh}, at $Ro=1.0$ in the northern hemisphere, shown at the same times as the previous figure: $t=0.48, 1.07, 1.54$.  The format is the same as in the earlier example figure, Fig.\ref{fig:sh_lines}.  }
    \label{fig:nh_lines}
\end{figure}

\begin{figure}[ht!]
    \centering
    \includegraphics[width=1.0\textwidth]{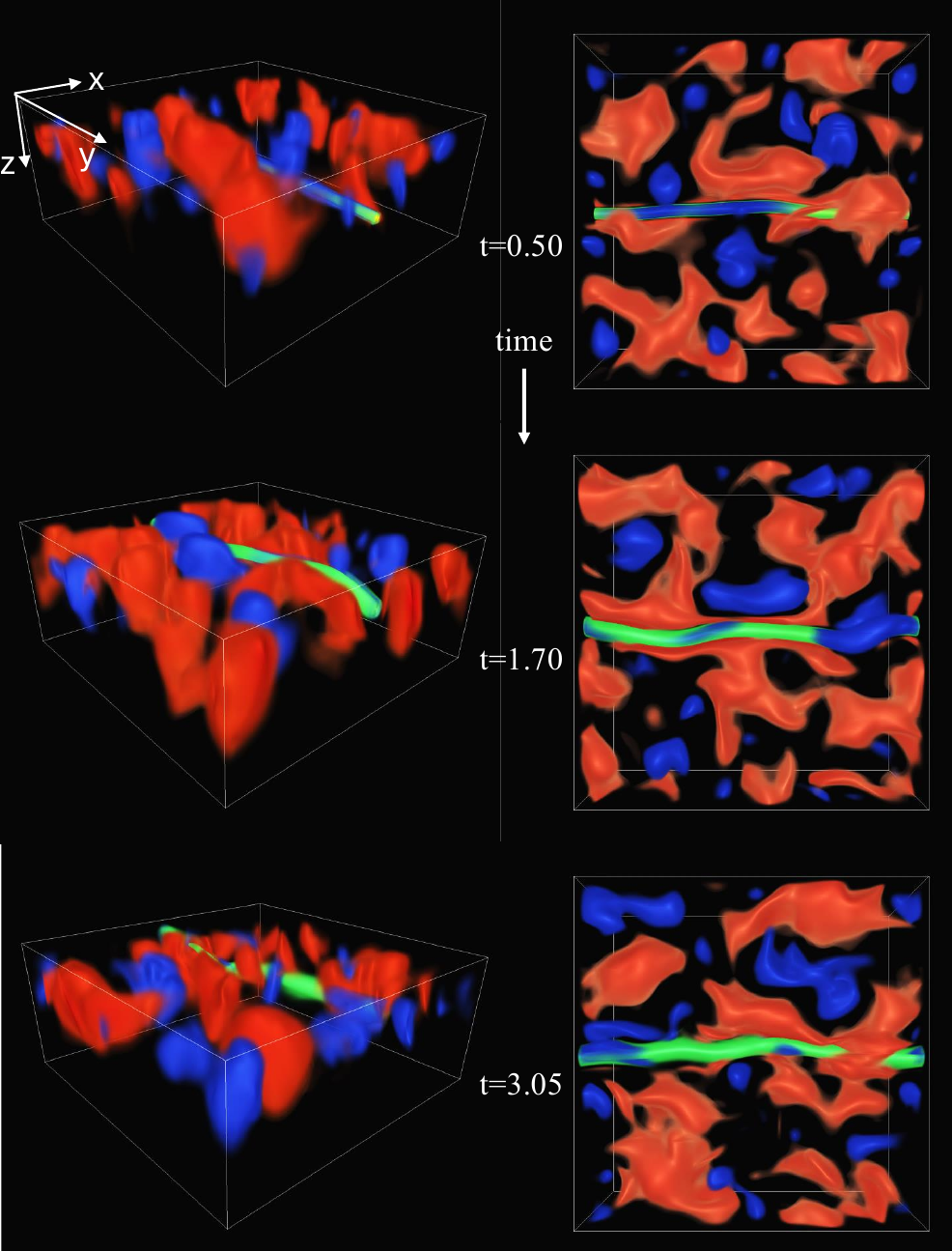}
    \caption{An example of a magnetic flux tube rising in the presence of rotating convection at $Ro=0.5$ in the northern hemisphere ($\phi=90^{\circ}$). The format is the same as the example figure, Fig. \ref{fig:sh}.   Shown are times $t=0.50, 1.70, 3.05$ (top to bottom).  }
    \label{fig:nh_Ro0p5}
\end{figure}

We now proceed to compare the evolution of $\mathcal{W}$ in convective simulations influenced by various levels of rotation, as measured by the Rossby number, $Ro$. Figure \ref{fig:W_north} plots $\mathcal{W}$ versus time for multiple realizations of rotating convection at two different Rossby numbers, $Ro=1.0$ and $Ro=0.5$, in panels (a) and (b) respectively.  All of these simulations are located in the northern hemisphere ($\phi=+90^{\circ}$) and performed at $Ra=4\times10^4$. Panel (c) of the figure compares the averages over the realizations at each $Ro$ value with those of the non-rotation and the non-convective (diffusive) cases.  We also complementarily visualize the effects of the flow on the flux tube from one of the realizations at each of the two $Ro$ values in Figures \ref{fig:nh}-\ref{fig:nh_Ro0p5}, in the same style as earlier figures.

From Figure \ref{fig:W_north}, for all realizations at both $Ro$, the overall evolution of the winding number, $\mathcal{W}$, is very similar to the non-rotating convective cases.  That is, the early evolution of $\mathcal{W}$ in each simulation follows that of the non-convective, purely-diffusive solutions as the flux tube first traverses the overshoot zone, but once the tube reaches the convective region and interacts with convective flows, $\mathcal{W}$ begins to drop significantly faster than the diffusive decay.  The individual realizations can deviate substantially from the average, and can further change drastically in time, but the average behavior (shown most clearly in panel $c$) is clearly a faster decay of $\mathcal{W}$.

Figure \ref{fig:nh} shows only one particular realization out of the many performed at $Ro=1.0$ but allows a small window into this type of observed behavior.  
By comparing this volume rendering (and Fig.~\ref{fig:nh_Ro0p5}) with the previous non-rotating version (Fig.~\ref{fig:noRot}), it can immediately be seen that the scale of the convection decreases with the increased influence of rotation, with smaller convective cells in the upper network, and more downflows initiated at the interstices of that network.  The flux tube therefore has a ``busier" convective regime through which to pass, as compared to the non-rotating convective case.
Despite this, the tube in this particular case seems to traverse the convective region without encountering much major disruption by any flows, even the strong downflows.  The flux tube is apparently only slightly deformed left-right and up-down by the buffeting effect of the convection. 
Figure \ref{fig:nh_lines} displays these dynamics by plotting several chosen field lines within the flux tube for this case at the same three times in the simulation as the three rows of Figure ~\ref{fig:nh}. The fieldlines at the current time are shown in red, and the fieldlines of the initial condition are shown in blue for reference.   Here, we can more clearly see the impact of the deformation of the tube on the field lines, elongating and unwinding in some regions and compressing and winding up in others.  Other realizations appear to have more dramatic large scale perturbations to the tube depending on the exact locations of strong downflows and upflows.  
It is difficult  visually to evaluate what the net effect on $\mathcal{W}$ might be, especially bearing in mind the net diffusive loss, of course. although one would definitely anticipate changes.  The deviations here do not seem dramatic and yet lead to a significant decrease in the winding number, as compared to the non-convective, purely-diffusive decay of a tube.
Ultimately we must rely on the veracity of the accounting of $\mathcal{W}$ portrayed in Figures \ref{fig:W_north}, which also notably averages over many more fieldlines per timestep than are shown in Figure ~\ref{fig:nh_lines}.  

Figure \ref{fig:W_north} in panel (c) therefore overlays the averages of $\mathcal{W}$ over the different realizations from the two cases at $Ro=1.0$ and $Ro=0.5$ in the northern hemisphere.  From this it is clear that the winding number, $\mathcal{W}$, appears to behave in essentially the same manner regardless of the rotation influence.  If a $\Sigma$-effect were to be the dominant effect in these dynamics, our simulations should be expected to demonstrate a bias whereby the flux tubes gain right-handed writhe (from the left-handed kinetic helicity of the rotationally-influenced convective flows), and consequently, by conservation of total helicity (writhe + twist), should compensate for this writhe via a gain in left-handed twist (equivalent to a loss of right-handed twist).  Our results should therefore exhibit an extra loss of the original right-handed twist of our flux tubes (from the initial condition) compared to the non-rotating case, with the amount of this extra loss increasing with increasing rotational influence on the flow (decreasing $Ro$).  We do not see this effect.  Figure \ref{fig:W_north}(c) does exhibit an accelerated loss of right-handed twist (i.e. a decrease of $\mathcal{W}$ versus time) compared to the non-convective, purely-diffusive case.  When compared to the non-rotating but convecting case though, the opposite of what we expect is seen. That is, the more rotationally-influenced cases (if anything) perhaps increase the initial right-handed twist in proportion to their increasing rotational influence. However, we believe that this gain is probably not significant and in fact the winding number evolution is essentially the same for the non-rotating and rotating cases,   This belief is due partially to the inherent large errors in the low population Monte Carlo averaging, and, more convincingly, to the further results for the southern hemisphere that follow in the next section.

Note that the tube rises somewhat more slowly in the presence of faster rotation, noticeable by a longer time trace for $\mathcal{W}$ for $Ro=0.5$ (that extends to $t_{\rm max} \sim 2.5$, as compared to $t_{\rm max} \sim 1.5$ for the $R=1.0$ case, where $t_{\rm max}$ is the time at which we cutoff the calculation as the flux tubes nears the top of the domain).  This fact, in conjunction with the smaller cell size for higher rotational influence, would tend to permit a more significant $\Sigma$-effect  (i.e., a larger ratio of rise time to turnover time, $\tau$, and more interaction with convective cells). Despite this benefit, the winding number, $\mathcal{W}$, seems to decay at a loss rate that is relatively independent of $Ro$, with the ultimate value of $\mathcal{W}$ depending only on length if time that the tube is resident in the convective turbulence.


\subsection{Evolution of \texorpdfstring{$\mathcal{W}$}{W} in the southern hemisphere (\texorpdfstring{$\phi=-90^\circ$)}{phi=-90}}

So far it has been difficult to distinguish any highly significant rotationally-influenced effect on the value of $\mathcal{W}$ in our rotating convection simulations. Perhaps a more definitive test is to reverse the rotation, or, place the domain in the southern hemisphere ($\phi=-90^\circ$) rather than the northern hemisphere ($\phi=+90^\circ$). If the $\Sigma$-effect were to be a dominant process, we would expect the opposite effect on $\mathcal{W}$ in the southern hemisphere to that in the northern hemisphere. That is, in the southern hemisphere, the chirality of the kinetic helicity of the convection should change sign, since the rotation has changed sense but upflows still diverge and spin down.  It then follows that the handedness of the magnetic writhe imposed on a flux tube by the kinetic helicity and the compensatory twist it then experiences should both also switch handedness. In our southern hemisphere simulations, following the $\Sigma$-effect theory then, we would expect our positively-twisted initial magnetic flux tube to gain further positive twist on interaction with the rotating convection (in proportion to the degree of rotational influence), rather than lose positive twist as should be the case for northern hemisphere simulations.

\begin{figure}[ht!]
   \centering
    \includegraphics[width=0.65\textwidth]{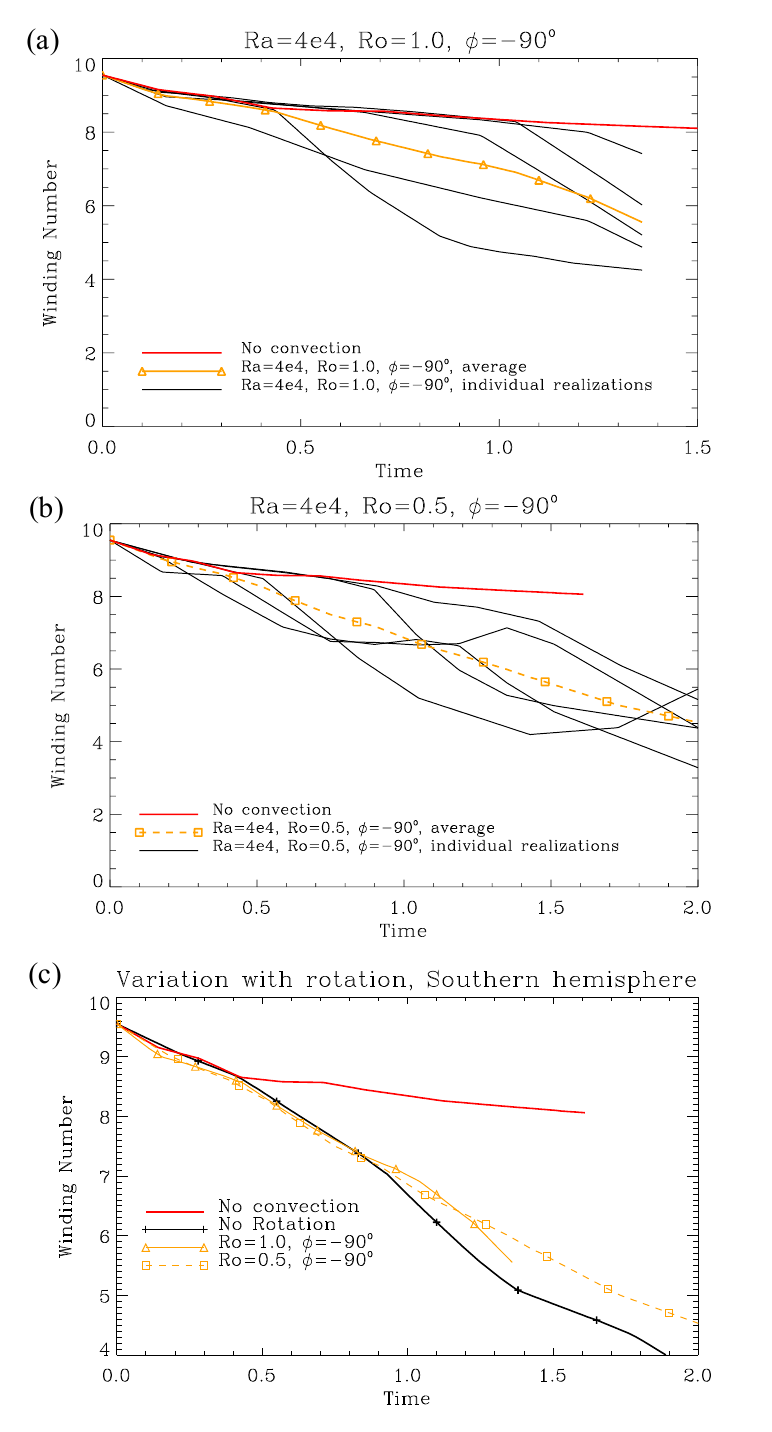}
    \caption{Evolution of $\mathcal{W}$ versus time for rotating convection at $Ra=4\times 10^4$ in the southern hemisphere ($\phi=-90^{\circ}$) for various rotation rates: $(a)$ $Ro=1.0$ and $(b)$ $Ro=0.5$. Each black line without a symbol corresponds to the evolution of an individual tube. The average over all realizations in each case is displayed thicker, in color, and with a symbol: orange with a triangle symbol for $Ro=1.0$ and dashed orange with squares for $Ro=0.5$. Panel (c) compares the average evolutions of $\mathcal{W}$ at each $Ro$ with the non-rotating case ($Ro=\infty$; black with + symbols) and the non-convective case (red).}
    \label{fig:W_south}
\end{figure}

Figure \ref{fig:W_south} shows results for the winding number, $\mathcal{W}$, from convection simulations placed in the southern hemisphere ($\phi=-90^{\circ}$) at two levels of rotational influence, $Ro=1.0$ and $Ro=0.5$, and fixed  $Ra=4\times10^4$.  This figure is the southern hemisphere counterpart of Figure \ref{fig:W_north}.  It can quickly be surmised that the non-convective, purely-diffusive decay (red line) is never counteracted in any noticeable fashion by the convective cases, nor are the rotating cases losing $\mathcal{W}$ more slowly than the non-rotating case.  Indeed, for both levels of rotational influence, the evolution of the average $\mathcal{W}$ over the realizations (see panel $c$) initially follows the non-convective, purely-diffusive solution (during the flux tube's rise through the overshoot zone), and then decreases more drastically as the tube interacts with the convection, but very closely follows the evolution of the non-rotating case without showing any significant difference due to rotation.  There are minor but probably insignificant deviations at later times, and more notably very little $Ro$ dependence.

\begin{figure}[ht!]
    \centering
    \includegraphics[width=1.0\textwidth]{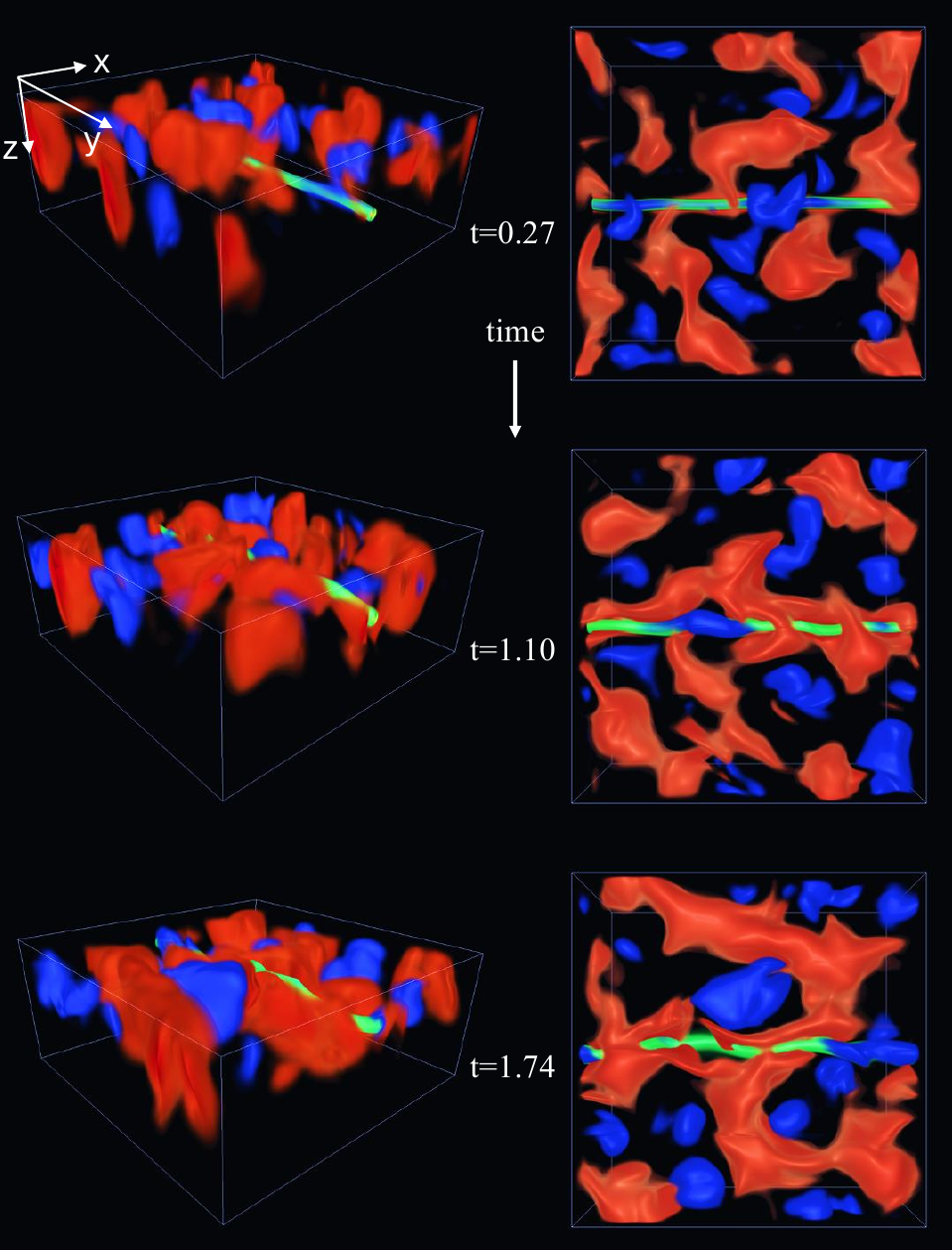}
    \caption{An example of a magnetic flux tube rising in the presence of rotating convection at $Ro=0.5$ in the southern hemisphere ($\phi=-90^{\circ}$). The format is the same as the example figure, Fig. \ref{fig:sh}.   Shown are times $t=0.27, 1.10, 1.74$ (top to bottom).   }
    \label{fig:sh_Ro0p5}
\end{figure}

The remarkably similar dynamics for these southern hemisphere cases are visualized for one realization from the $Ro=1$ simulation in the earlier Figures \ref{fig:sh} and \ref{fig:sh_lines} (used as examples earlier in the text), plus Figure \ref{fig:sh_Ro0p5} which shows one realization for $Ro=0.5$. 
In Figure \ref{fig:sh_lines}, it can be seen that the flux tube becomes considerably distorted, and we observe, especially at the later times, that two large portions of the tube have risen much further than the rest of the tube.  This is perhaps easiest to see in the black projection of the centerline of the rising tube on the right-hand wall. Figure \ref{fig:sh} shows the accompanying volume renderings for direct comparison with Figure \ref{fig:nh}.  Here, 
we can see both strong upflows (blue) pushing portions of the flux tube field upwards (at around one-third and two-thirds of the length of the tube) and strong downflows (red) in between holding other portions of the tube (the middle and ends) down. These competing flows swiftly deform the tube quite drastically (mostly in the vertical direction) and cause field line winding to be significantly affected.  However, as mentioned before, visual assessment of the net result is difficult and we should instead rely on the computed accounting of $\mathcal{W}$.  Here, this reveals that there is always only a fairly universal rapid reduction in value, representing net unwinding, and no clear variance in the rate of this unwinding with variation in the rotational influence.  Furthermore, we have now shown that there seems to be little or no dependence in the rate of loss of $\mathcal{W}$ even with a switch in the sign of the helical nature of the convection as is the case when the simulations are switched between the northern and southern hemispheres.


\subsection{Evolution of \texorpdfstring{$\mathcal{W}$}{W} at increased \texorpdfstring{$Ra$}{Ra}}

Despite the coarse nature of these explorations, it appears that, in all cases so far, the deformations of the magnetic structures by the larger-scale convective upflows and downflows seem to dominate over any smaller-scale turbulent kinetic helicity effect, and the trend is always to unwind the twist of the structure at a rate faster than that of diffusion independent of rotational influence and hemisphere. One reason we might be unable to detect any rotational influence is that the convection in these simulations may not have a sufficiently wide range of scale separation between the turbulent scale and the magnetic structure scale to be in a regime consistent with the assumptions of $\Sigma$-effect theory. Looking at the previously shown volume renderings of convective velocities again, one would get the impression that the convection is dominated by large-scale features and not by small-scale turbulent features much smaller than the tube length scale, although part of this is due to the vizualization techniques chosen where we highlight strong values. 
As a caveat, it should also be noted that it is the larger-scale convective features that might be expected to contain the sense of kinetic helicity that the $\Sigma$-effect demands. Regardless, in an attempt to increase convective vigor and its associated range of scales, and perhaps subsequently achieve dynamics closer to those of the $\Sigma$-effect models, we have performed additional simulations at $Ro=0.5$ (our strongest level of rotational influence) in the northern hemisphere where the Rayleigh number is increased five-fold to $Ra=2\times 10^5$.

\begin{figure}[ht!]
    \centering
    \includegraphics[width=1.0\textwidth]{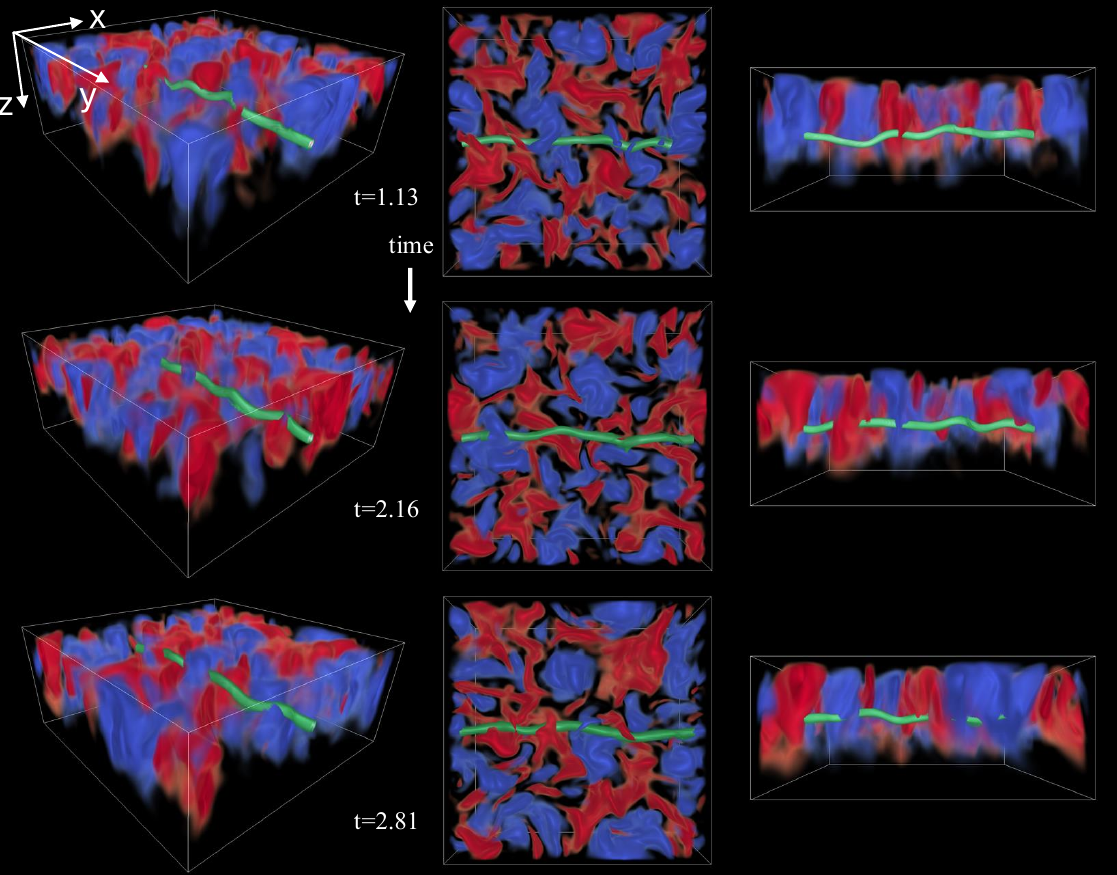}
    \caption{An example of a magnetic flux tube rising in the presence of rotating convection at higher $Ra=2\times10^5$ than previous simulations, and at $Ro=0.5$ in the northern hemisphere ($\phi=90^{\circ}$). The format is the same as the example figure, Fig. \ref{fig:sh} except that a third panel showing a side view is added to each time shown in a row.  Shown are times $t=1.13, 2.16, 2.81$ (top to bottom). In this higher $Ra$ case, the tube rises only slightly and then sits in place near the bottom of the convective region allowing much greater scope for the stronger turbulence to act. }
    \label{fig:hira}
\end{figure}

\begin{figure}[ht!]
    \centering
    \includegraphics[width=1.0\textwidth]{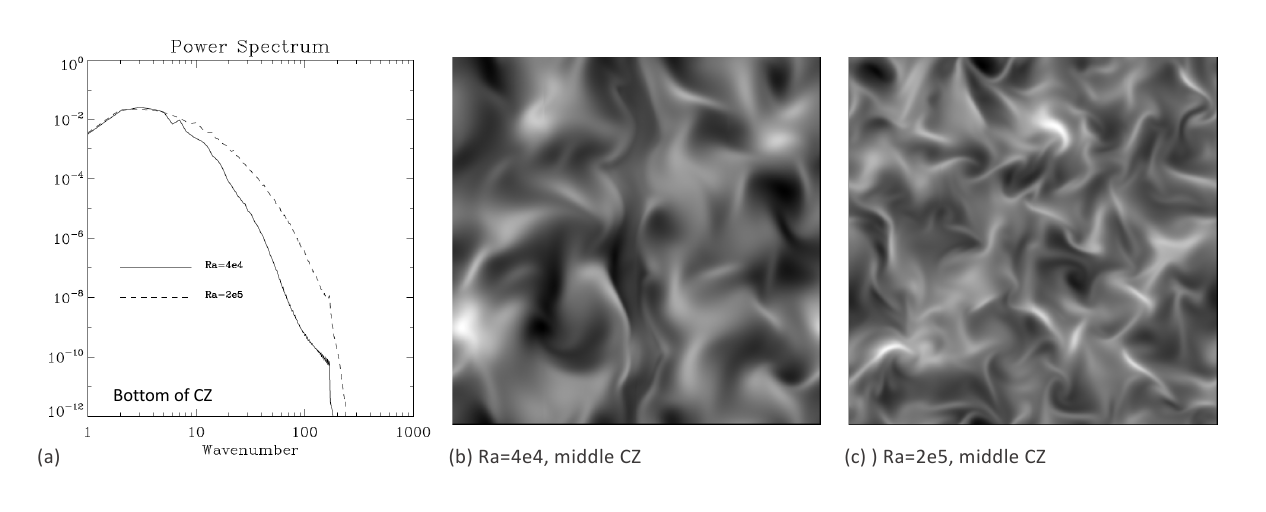}
    \caption{Characteristics of the high Rayleigh number case at $Ra=2\times10^5$.  (a) The horizontal power spectra taken at $z=1.0$, the bottom of the convection zone, for the two cases $Ra=4\times10^4$ and $Ra=2\times 10^5$.  (b)-(c) Grey scale images of the vertical velocity, $w$ at $z=0.5$, the midpoint of the convection zone, for each of the cases at $Ra=4\times10^4$ and $Ra=2\times10^5$ rspectively.  Overall, it is clear that the higher Rayleigh number case is more turbulent, with more power in smaller scales.  }
    \label{fig:powerspec}
\end{figure}

Figure \ref{fig:hira} shows volume renderings for one realization at this new higher Rayleigh number in a similar manner to that used for previous figures (such as Figure \ref{fig:nh}).  It can immediately be seen that the scale of the convection has been reduced as compared to the equivalent case at $Ra=4\times10^4$ 
and that the flows look significantly more turbulent, as desired. Figure \ref{fig:powerspec} further confirms this, comparing details of the standard and high $Ra$ cases.  Slices of the vertical velocity, $w$, at the midpoint ($z=0.5$) of the convection zone (panels $b$ and $c$) show the differences in scales between the two cases more clearly than the volume renderings.  Furthermore, in panel $a$, the power spectrum for each case is shown,  calculated at the bottom of the convection zone where the magnetic flux tube spends more of its time (especially in the higher $Ra$ case, as will be discussed shortly).  This figure confirms more quantatively that the higher $Ra$ case has more power at smaller scales and is therefore overall more turbulent.

\begin{figure}[ht!]
    \centering
    \includegraphics[width=1.0\textwidth]{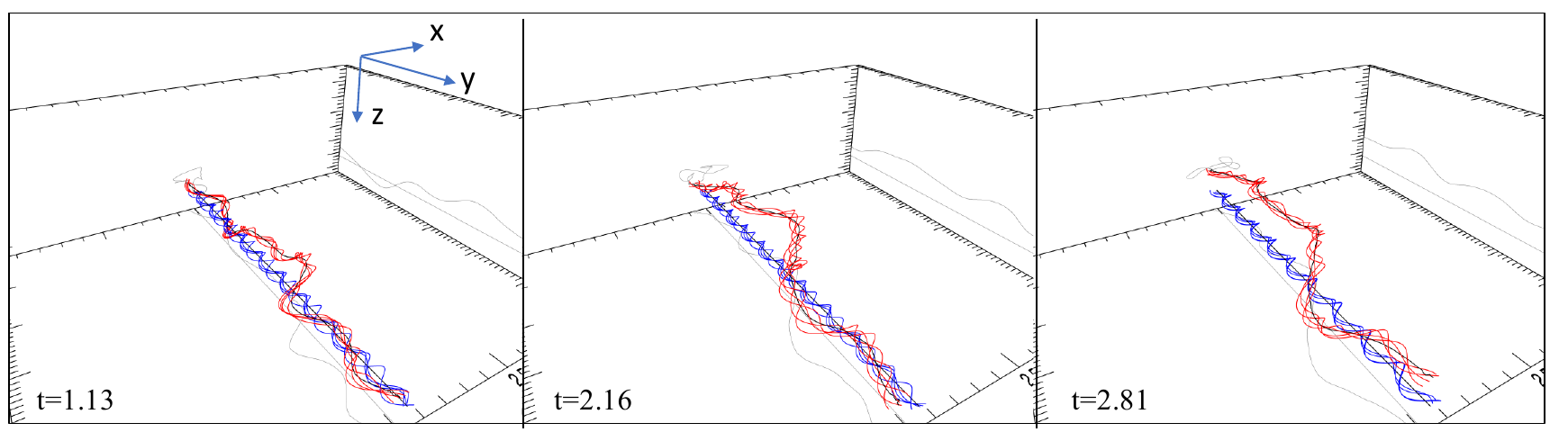}
    \caption{Illustration of the field line contortion in the magnetic flux sdtructure for the case shown in Fig.\ref{fig:hira} at a higher $Ra=2\times10^5$ and $Ro=0.5$ in the northern hemisphere.  The panels show the same times as in the previous figure:$t=1.13, 2.16, 2.81$.  The format is the same as in the earlier example figure, Fig.\ref{fig:sh_lines}. }
    \label{fig:hira_lines}
\end{figure}

\begin{figure}[ht!]
    \centering
    \includegraphics[width=1.0\textwidth]{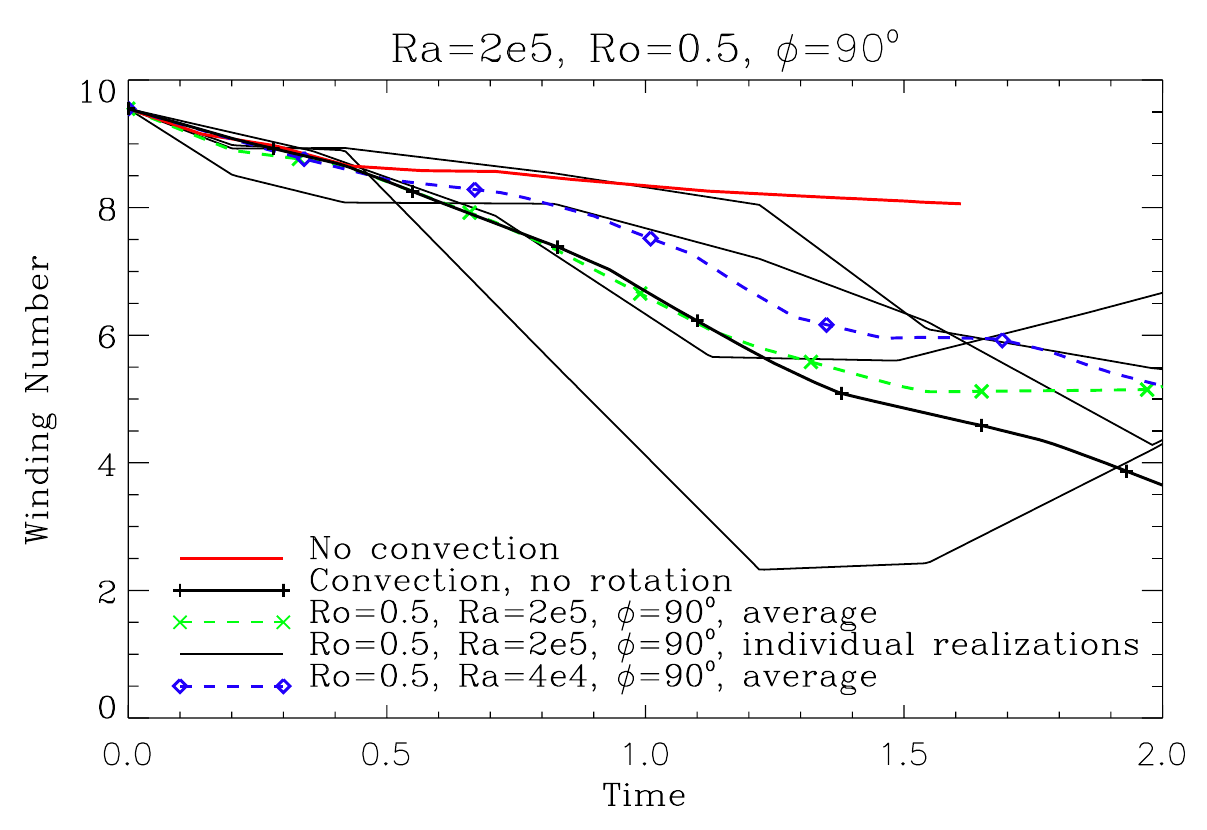}
    \caption{Evolution of $\mathcal{W}$ versus time. Each black line corresponds to an individual tube evolving in rotating convection at $Ro=0.5$, $Ra=2\times 10^5$, $\phi=90^\circ$, with the average displayed in green with x symbols. The non-convective diffusive decay is shown in red for comparison.}
    \label{fig:W_Ro5en1_lrgRa}
\end{figure}

Figure \ref{fig:hira_lines} shows the contortion of some of the field lines in the magnetic flux structure  during the evolution of the simulation at high $Ra$.  The times shown correspond to the same times as shown in the volume renderings in Figure \ref{fig:hira}. Significant contortion can be seen as might be expected from buffeting by the smaller scale turbulence of this case.  More quantitatively, the evolution of $\mathcal{W}$ for these higher $Ra$, more turbulent simulations are plotted in Figure \ref{fig:W_Ro5en1_lrgRa}. Overall, we again observe the now familiar behavior of this measure, i.e.~an initial evolution that follows the decay of $\mathcal{W}$ in the non-convective, purely-diffuse simulations but which then subsequently decreases more rapidly. The variance between the evolutionary tracks for the (admittedly small number of) realizations here seems to be more dramatic, as perhaps might be expected with higher levels of turbulence implemented in this case. The time traces for individual realizations here also sometimes exhibit quite strong deviations to faster loss of $\mathcal{W}$, but at other times show substantial increases.  Such deviations unsurprisingly point to the necessity of averaging over a greater number of realizations as the degree of turbulence increases, a laudible practice that is sadly restricted by the computational expense. With the current number of simulations, we see that the average loss of $\mathcal{W}$ appears to be more rapid for the higher $Ra$ (green dashed line with x symbols) than the lower $Ra$ (blue dashed line with diamond symbols).  This initially appears consistent with the presence of a more effective $\Sigma$-effect at higher $Ra$, but unfortunately the rate of loss of $\mathcal{W}$ for both cases is still not faster than the non-rotating case (thick black solid line with + symbols), which is inconsistent with the $\Sigma$ model.  If we are optimistically trusting the averages here, it is conceivable that even higher $Ra$ might enhance the rate of loss of $\mathcal{W}$ to a more rapid value than that of the non-rotating case, but we have not explicitly seen this here, and there are some reasons why this might not be likely, as will be expounded in the discussion. 

\begin{figure}[ht!]
    \centering
    \includegraphics[width=1.0\textwidth]{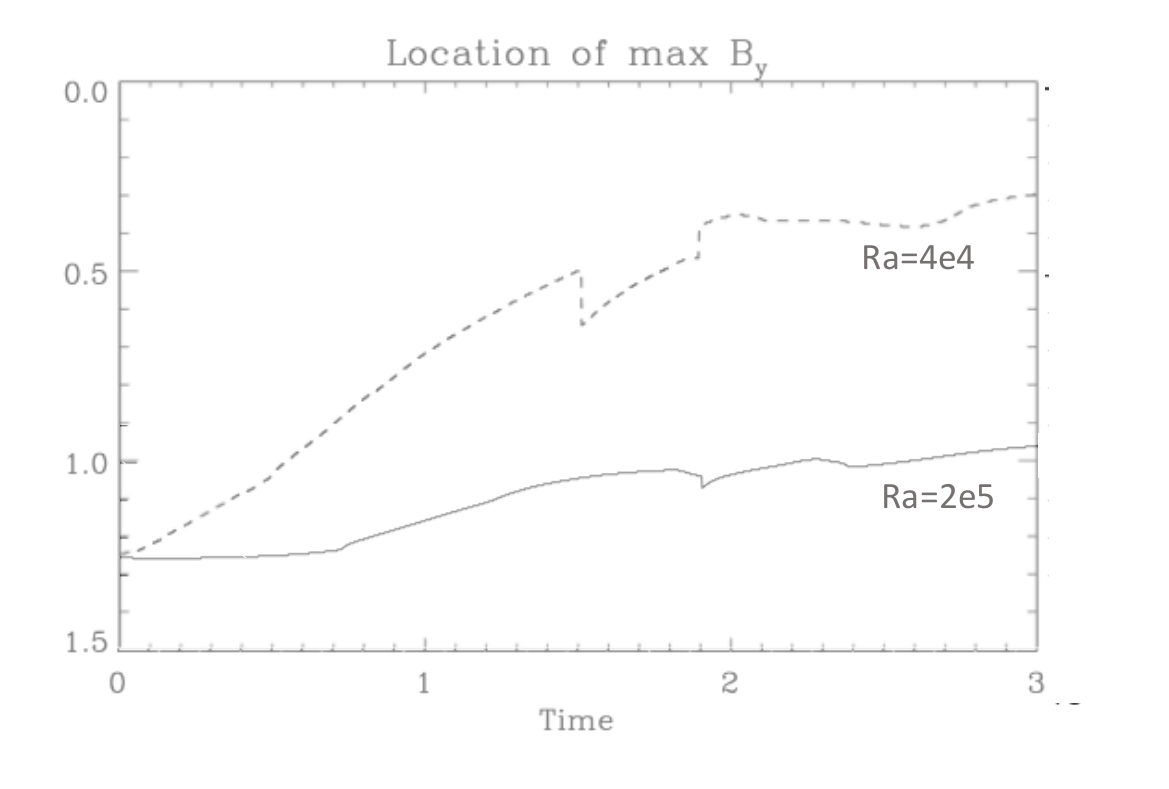}
    \caption{Location of the maximum $B_y$ in two simulations at the same Rossby number, $Ro=0.5$, in the northern hemisphere but at different Rayleigh numbers, one at $Ra=4\times10^4$ and the other at $Ra=2\times 10^5$.  This indicates the different rates of rise of the magnetic structures.}
    \label{fig:b2vst}
\end{figure}

One difference worth noting in this case is that the flux tube does not rise quickly to the top of the domain as before, but rather only rises slightly in the beginning and then sits near the base of the convective region. This can be seen in the volume renderings of Figure \ref{fig:hira}, and in Figure \ref{fig:b2vst}, where we plot the vertical location of the maximum magnetic energy versus time, anticipating that this gives us a reasonable (if potentially discontinuous) sense of the location of the tube as a whole.  As can be seen in the latter plot, by this measure the tube  in the high $Ra$ simulation has only risen slightly into the lower convection zone ($z \sim 1$) by time $t=3.0$, a time at which most tubes in previous simulations have risen completely to the top boundary of the domain ($z=0$).  This slower rise occurs because, when increasing $Ra$ at fixed $Ro$, the value of $Ta$ must also correspondingly increase, leading to stronger horizontal mixing of the motions by rotational effects, and subsequently a reduction in the rise rate of the magnetic structure.  This slower rise could also have contributed to the more significant $\Sigma$-effect style behavior in this case, highlighting the difficulties of directly comparing cases at different parameters as mentioned earlier.
In order to counter this effect and to increase the rise speed to make it commensurate with the other simulations, we could increase the magnetic field strength (governed by $\alpha_m$). However, at the convective parameters we are using, the ability to do this is limited, since too large of an increase in $\alpha_m$ will result in a negative density deficit within the tube when correcting the thermodynamics of the tube for the presence of the magnetic field. An interesting conundrum therefore arises (in these simulations or in the Sun) as to how to achieve sufficiently strong magnetic fields in order to generate a rise through the convection zone while at the same time providing sufficiently strong rotational constraints to rotationally influence the small scales of highly turbulent convection.


\section{Discussion}\label{sec:disc}

Our goal with this paper has been to examine a more realistic implementation of the ideas inherent in the $\Sigma$-effect model of \cite{longcope1998flux}.  The latter model stands as perhaps the most widely cited explanation for the origin of the solar hemispheric helicity rules (SHHR) at the current time.  
The SHHR \cite[see e.g.][]{pevtsov1995latitudinal} is a weak observational rule  
that describes a correlation between the hemisphere that an active region emerges in and the handedness of its magnetic helicity (or more correctly, the radial component of the magnetic fieldline twist, a component of the total current helicity).  
The original model in \cite{longcope1998flux} provided a very elegant but highly idealized conceptual model for how an untwisted flux tube 
may become twisted when acted upon by helical turbulence.  Conceptually, the main idea is that the kinetic helicity of the rotational turbulence could impart a systematic writhe to the axis of the flux tube 
and therefore induce an opposing systematic twist within the flux tube as it attempts to conserve the total helicity (writhe plus twist).   
With a specific but common  assumption about the handedness of the kinetic helicity of solar convection under the influence of the rotation of the star, this model leads naturally to an expectation of the chirality of the twist in magnetic structures that is commensurate with that observed in the SHHR.
The model used in \cite{longcope1998flux} to explore this conceptual idea is highly simplified in the sense that, firstly, the flux tube is modeled as a thin flux tube \citep[i.e.~simply a one-dimensional curve that moves in space and time with assigned dynamical properties such as buoyancy, drag and twist, but no cross-sectional area; see e.g.][]{Spruit:1981}, and, secondly, the convection through which the structure rises is incorporated solely as an influence on the writhe of the tube axis due to a helical turbulence model (via the eponymous $\Sigma$ term).  
Our aim here has been to investigate a macro-scale, less idealized model of the same underlying dynamics. 
To this end, we numerically simulate the evolution of an artificial (imposed), but now fully three-dimensional magnetic structure (a finite cross-sectional ``magnetic flux tube'') rising by magnetic buoyancy through a domain containing fully-resolved, three-dimensional, rotationally-influenced compressible convective turbulence.  Our ultimate goal has been to see if we could recover the direct relationship between the kinetic helicity of the convection and the resulting chirality of the magnetic structures after transport through the convection, which is the essence of the $\Sigma$-effect.

Our three-dimensional magnetic flux tubes have the advantage that we can directly examine the evolution of the magnetic field lines within our flux structures.  Our most effective tool is the winding number ($\mathcal{W}$; see Equations \ref{eqn_wi_1}-\ref{eqn_windingnumber}) that quantifies the degree of twist of field lines around the axis of the magnetic structure at any time during the evolution. 
If the $\Sigma$-effect were to be a dominant dynamical effect, there would be a direct relationship between the evolution of the twist of a magnetic structure, as measured by the change in $\mathcal{W}$, and the helical nature of the turbulence through which it travels.  The kinetic helicity of the convection is determined by the Rossby number, $Ro$, and the  hemisphere in which the convection operates, set via the parameter $\phi$.
We therefore performed  fully three-dimensional simulations varying these parameters while tracking changes in the winding number $\mathcal{W}$ over time as the magnetic flux tube rises from the deeper radiative interior of the domain into and through the convective region.  For calibration, we compare each evolution of $\mathcal{W}$ with that of cases without convection, where the evolution is purely diffusive, and with that of cases where there is convection but no rotational influence. 
For any particular parameter set, a number of realizations were performed in an effort to provide a minimal Monte Carlo approach that enables some averaging over the significant fluctuations introduced by spatial and temporal  variations in the initial conditions.  

\begin{figure}[ht!]
    \centering
    \includegraphics[width=1.0\textwidth]{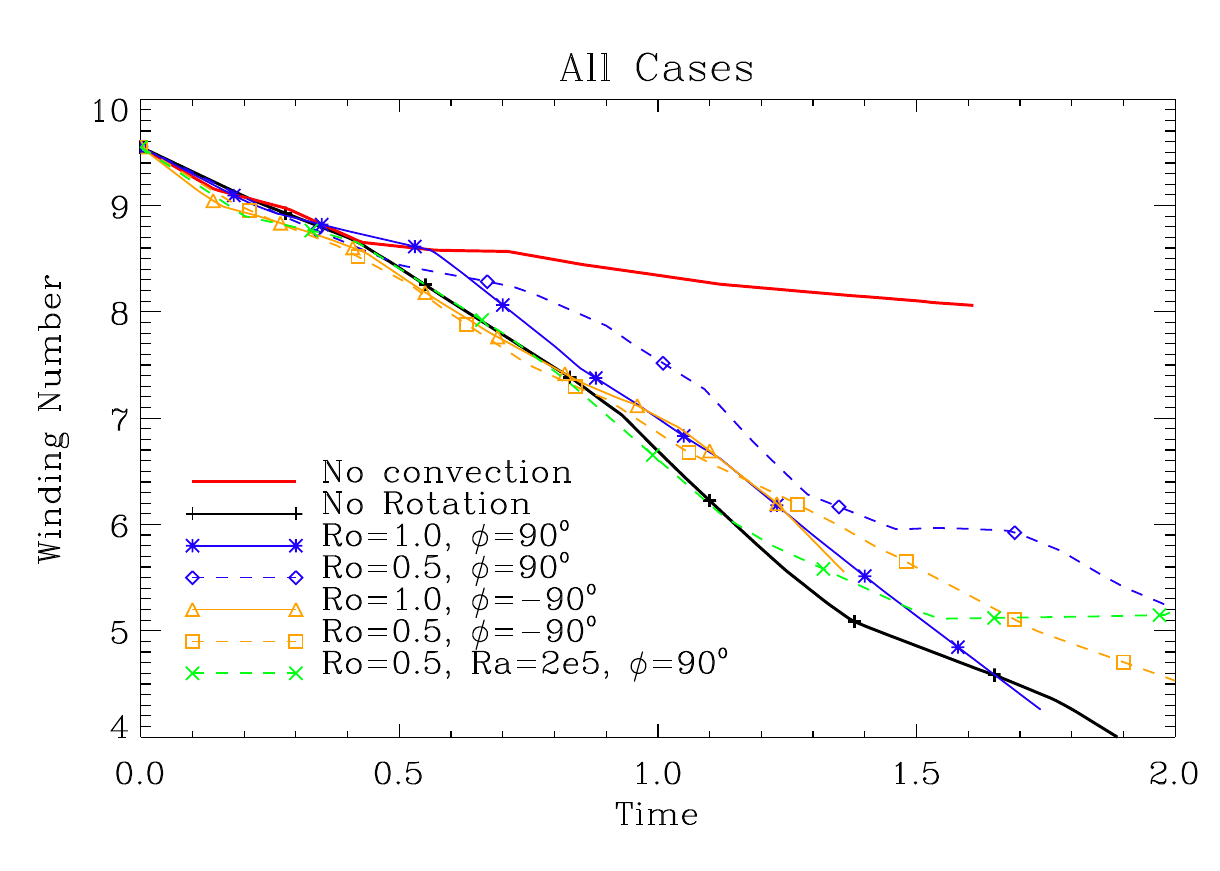}
    \caption{Evolution of $\mathcal{W}$ versus time displaying the averages of each parameter set.}
    \label{fig:W_avg}
\end{figure}

All of our results are collected and summarized in Figure \ref{fig:W_avg}. This figure shows the evolution of $\mathcal{W}$ averaged over all the Monte Carlo realizations at each parameter set that we have computed.  The curves plotted represent the average behavior of a number of Rossby number values ($Ro=\infty,1.0,0.5$) in both hemispheres ($\phi=\pm 90^{\circ}$). Overall, we found no clear indication of a $\Sigma$-like effect in our simulations. That is, we could not discern the expected Rossby number dependence of $\mathcal{W}$ over time, nor could we even see a clear systematic distinction between simulations performed in opposite hemispheres. Instead, in all cases with any sort of convection, we observed a fairly universal  decrease in $\mathcal{W}$ that was substantially faster than the diffusive decrease found in the non-convection results. We therefore tentatively conclude that we are not witnessing a rotational influence such as that predicted by the $\Sigma$-effect, but rather an effect that, in mean field terminology, is more akin to a ``turbulent'' or ``enhanced'' diffusion associated with the random turbulent interactions between the convective flows and the magnetic structure. 
{
This lack of evidence for the $\Sigma$-effect was also found observationally in the work of \cite{Liu2024}, although the connection of the two works has some major caveats, notably that the observational measures of kinetic and magnetic helicity could only be accessed very close to the surface and not in the deeper interior as simulated here.
}

Perhaps the most obvious reaction to this discovery is to question whether the conditions required for the $\Sigma$-effect were actually met. In particular, one might wonder if the convection simulated were supercritical enough to have a substantial range of small-scale helical turbulence that could influence the magnetic flux structure.  With this idea in mind, we performed simulations at higher Rayleigh number ($Ra=2\times10^5$), that had significantly more power in the smaller scales.  However, the results from these cases were not substantially distinguishable from the previous results, and again do not exhibit the expected trend in the evolution of the twist with rotational influence expected from the $\Sigma$-effect.

There are many things one could criticize in our simulations, that are worth mentioning here.  First of all, we do note that different realizations of convection at the exact same parameters produce substantially different changes in $\mathcal{W}$ and that we have only averaged over a small set of realizations in our Monte Carlo suite for each parameter set. We already know from observations that the $\Sigma$-effect bias is not a universal rule, being obeyed only 60\% of the time.  We therefore might need to average over a much larger sample size in order to observe this level of bias. Our statistics here are admittedly not great and the error bars are substantial, but this is due to the substantial computational expense of such simulations. 

Compounding this, it is difficult to measure an accurate and consistent value for $\mathcal{W}$ with turbulent and diffusive effects present. For example, identification of what constitutes the magnetic flux structure (``flux tube'') and its fieldlines is not a unique nor precise process. As the flux tube interacts with convection, magnetic field lines can be advected rapidly by the convective flows and occasionally end up being directed out of the domain or being diverted far from the bulk of the field lines that we take to constitute the ``flux tube''.  In these cases where a tube becomes substantially fragmented, it can be difficult to say whether the field lines should be considered part of the flux tube and be included in the calculation of $\mathcal{W}$ or not.  Furthermore, defining a centerline about which the winding number $\mathcal{}W$ is calculated, and even defining where the tube is (which we do by thresholding) is somewhat subjective.  More technically, if a tube loses its $y$-directional monotonicity, the calculation of the winding becomes very complex.  

Furthermore, it should also be noted that it is hard to guarantee direct comparability across our suite of simulations when altering parameters. Given that we wish to measure the effects of the transfer of kinetic helicity of the rotating convection to the magnetic structure, we ideally want to make sure that the convection has the same opportunity to impart its properties onto the magnetic field in all convective cases.  
To do this, we would have to alter more than just the rotation. For instance, the buoyant rise speed of the flux tube is dependent not only on the strength of the field, but also on the influence of rotation; rise speeds increases with stronger field strength, $\alpha_m$, but also decrease with stronger rotational influence, $Ro$. Therefore, to have similar rise times for a varying rotation rate, we should perhaps consider compensating by varying the initial field strength. Also, in order for the flux tube to remain coherent as it rises, the strength and twist of the tube must be sufficient to hold it together against the shredding effects of its wake and the vigorous convective flows. In terms of parameters, the tendency of a magnetic tube to remain coherent during a buoyant rise increases with the strength of the field, $\alpha_m$, and with larger twist, $q$, but decreases with stronger convective vigor, $Ra/Ra_{crit}$ (and where further  $Ra_{crit}$ depends on the rotational influence via the Taylor number, $Ta$).  With all these potential parametric variations in the rise characteristics, it is difficult to find a path through parameter space that guarantees clean comparability of the solutions.  
{We note in particular that we chose a larger $q$ than the minimum necessary for survival of the transit and this may have made smaller perturbations in twist due to any $\Sigma$-effect harder to detect.}

\begin{figure}[ht!]
    \centering \includegraphics[width=0.49\textwidth]{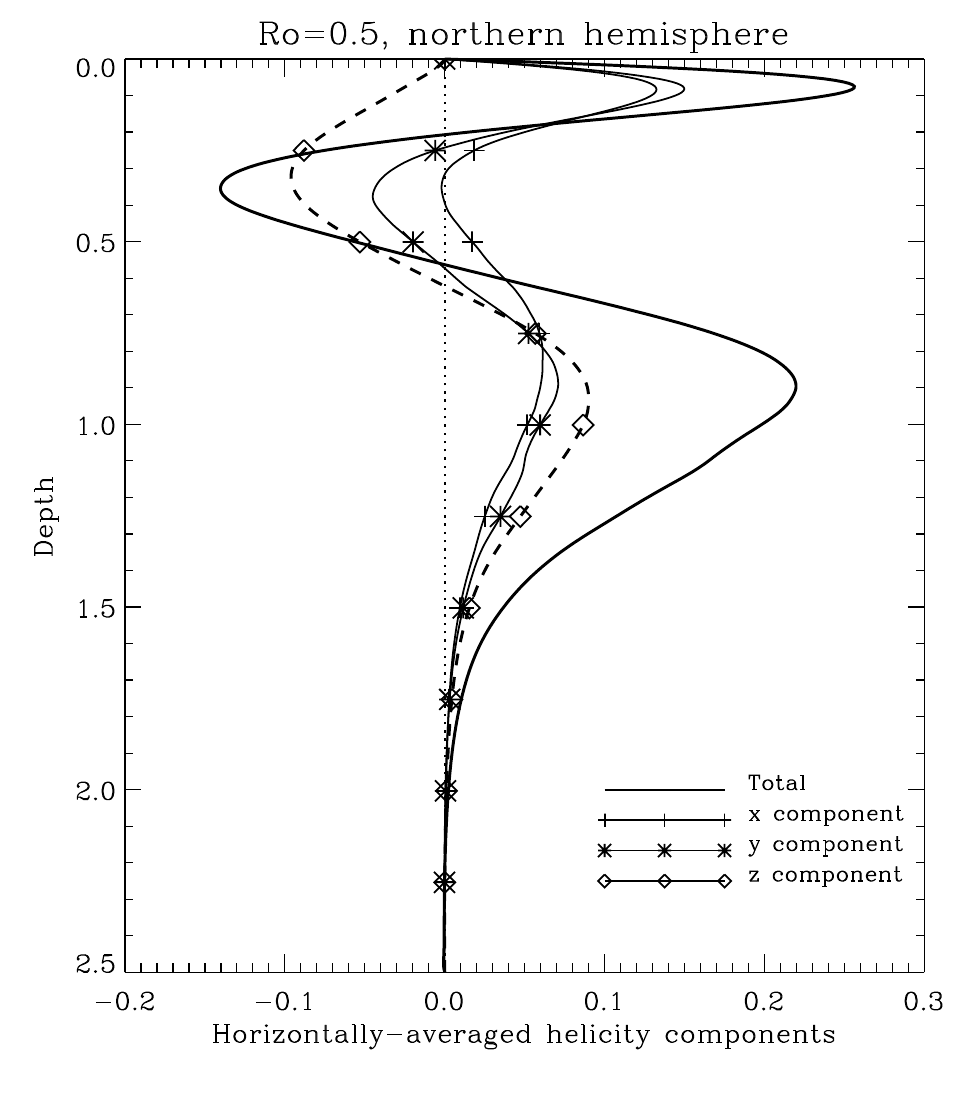} \includegraphics[width=0.49\textwidth]{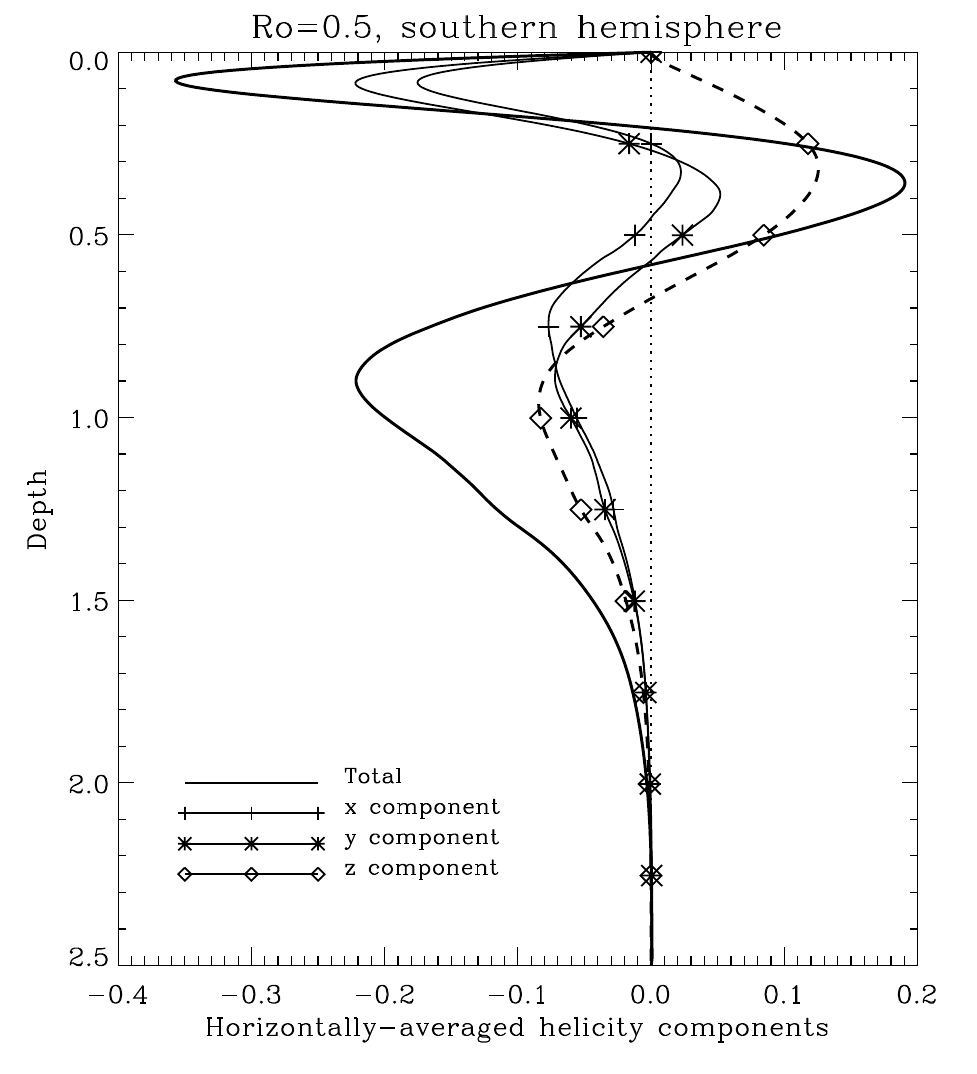}
    \caption{Horizontally- and time- averaged components of the kinetic helicity as a function of depth for (a) the northern hemisphere, and (b) the southern hemisphere. }
    \label{fig:kin_hel}
\end{figure}

Despite such shortcomings of our methodology, we believe that there are also some potentially viable theoretical reasons for the major result that we have found, i.e.~the lack of a distinguishable $\Sigma$-effect.
Perhaps most concretely, it should be noted that the study of \cite{longcope1998flux} uses an initial model for the helical turbulence (``Model H") that contains a specific single-signed chirality imposed throughout the buoyant evolution of the tube, namely negative kinetic helicity to represent the northern hemisphere or positive kinetic helicity for the southern hemisphere. 
This characterizes the general idea that, in compressible convection, rising fluid generally expands and spins down, whereas downflows contract and spin up. However, as astutely recognized in the original paper, this is not always the case in full simulations of rotating convection; the vertical profile of horizontally-averaged kinetic helicity can indeed switch signs with depth  \citep[see, e.g.,][]{brummell1998turbulent}.  This is often interpreted as ``splashing'' of downflows against a lower boundary in the simulations, thereby enforcing divergence of downward moving fluid, leading to the opposite kinetic helicity than that expected in general. The original $\Sigma$-effect paper accounted for this by introducing a second turbulence model (``Model B") that had a layer of reversed kinetic helicity, but the authors found no significant change in their results.
The lack of a single-signed kinetic helicity could then potentially be attributed to the unphysical inpenetrable boundaries in pure convection Cartesian box simulations.  However, the sign switch also occurs in our simulations here, which notably do not possess such a rigid lower boundary to the convection zone. To illustrate this, we plot in Figure \ref{fig:kin_hel} the time- and horizontally-averaged kinetic helicity versus depth for strong rotational influence ($Ro=0.5$) in both hemispheres.  The individual directional elements that make up the total helicity density are also shown.  The argument for the expected kinetic chirality is usually derived from the vertical element $w \cdot \omega_z$, where $w$ is the vertical velocity and $\omega_z$ is the vertical component of vorticity.  It can be seen in the plots that this component (the dashed line) does indeed possess the expected chirality (negative in the northern hemisphere, positive in the southern) in the upper 60-70\% of the convection zone.  However, deeper down, all elements, including the vertical element, reverse sign, to combine into a total helicity that has the opposite sign than that which is anticipated for the $\Sigma$-effect.  This behavior persists to the base of the overshoot zone where motions peter out ($z \sim 2$) .  The near-surface of the domain has more complicated behavior than perhaps expected, with the horizontal components flipping the sign of the total helicity there too.
Therefore, in our simulations where the rising flux tube is initiated in the overshoot zone, the magnetic structure only experiences the expected sign of total helicity as it evolves through a small vertical window between $z \sim 0.5$ and $z \sim 0.2$, and experiences the opposite sense in all other locations.  This effect could play a substantial role in explaining why our evolving structures appear not to acquire a specific chirality.  


With all this in mind, it is perhaps worth considering what one might need to do to give the $\Sigma$-effect a better chance of operating successfully.  Solar convection operates at extremely high Rayleigh and Reynolds numbers.  Our highest $Ra$ simulations performed ($Ra=2\times10^5$) were a simple attempt to move in the correct direction in parameter space.  These cases indeed allowed more buffeting of the magnetic structure by small-scale helical turbulence.  However, the continued lack of a clear chirality bias in the results returned by these cases perhaps reveals an inherent weakness in the conceptual $\Sigma$-effect model.  That is, as we extend the spectrum of the convection to smaller scales at higher $Ra$, it becomes increasingly more difficult to have those scales remain rotationally-influenced (and therefore helical as is required for the $\Sigma$-effect) without decreasing the $Ro$ of the simulation beyond reasonable values for the solar context.  The Sun is not a particularly fast rotator and only at relatively large scales (larger than the supergranular scale at the observable surface) are they directly rotationally influenced.  However, we must acknowledge that little is known about the turbulence of the deeper interior and, in particular, its typical scales and turnover times.  It is also further possible that small interior scales could be indirectly organized and influenced by rotationally-constrained larger flow structures.  Therefore at this juncture, it is not possible to completely rule out a $\Sigma$-like effect, but its operation becomes questionable.

On a perhaps even more pessimistic note, overall it becomes difficult to imagine how magnetic structures can emerge from the deep interior as one tends towards more solar-like parameters in simulations if such processes are required to be occurring.  Increasing the degree of turbulence towards more solar levels also increases the downward transport of large-scale magnetic fields \citep[``magnetic pumping''; see e.g.][]{tobias2001transport}.  Raising the rotational influence in order to allow the small scales the chance of operating a $\Sigma$-effect further reduces the rate of rise of magnetic structures (see e.g. Fig. \ref{fig:b2vst}).  It appears that more realistic parameter pathways that include the $\Sigma$-effect sadly lead to a more difficult transit through the convection zone for buoyantly rising magnetic structures. These restrictions could only be mitigated by an origination process that leads to much stronger structures somehow or a very precise balance of all the dynamical ingredients.

\begin{acknowledgments}
We thank B. Manek and Y. Liu for helpful discussions. This work was supported by the COFFIES DRIVE Science Center (NASA grant 80NSSC22M0162). Computational resources were provided by the Texas Advanced Computing Center (TACC) at The University of Texas at Austin via the NSF ACCESS program.
\end{acknowledgments}


\bibliography{winding}{}
\bibliographystyle{aasjournal}



\end{document}